\documentclass[useAMS,usenatbib]{mn2e}

\usepackage{graphicx}
\usepackage{epsfig}
\usepackage{times}
\usepackage{natbib}

\def\bm#1{{\hbox{\boldmath $#1$\unboldmath}}}

\def\vec#1{\bm{#1}}

\def\bm#1{{\hbox{\boldmath $#1$\unboldmath}}}
\def\vec#1{\bm{#1}}

\newcommand{\be}{\begin{equation}}
\newcommand{\ee}{\end{equation}}
\newcommand{\bea}{\begin{eqnarray}}
\newcommand{\eea}{\end{eqnarray}}

\def\eg{{\rm e.g.}}
\def\ie{{\rm i.e.$\,$}}

\newcommand{\Tobs}{{T}}
\newcommand{\Tcmb}{{T}}
\newcommand{\Tred}{{T_{\rm uncorr}^{\rm rec}}}
\newcommand{\Tknown}{{T_{\rm corr}^{\rm rec}}}
\newcommand{\Tknownreal}{{T_{\rm corr}}}
\newcommand{\Tredreal}{{T_{\rm uncorr}}}

\newcommand{\Pknown}{{P_{\rm corr}}}
\newcommand{\Pred}{{P_{\rm uncorr}^{\rm rec}}}
\newcommand{\Predreal}{{P_{\rm uncorr}}}
\newcommand{\Predraw}{{P_{\rm uncorr}^{\rm raw}}}
\newcommand{\Prec}{{P_{\rm cmb}^{\rm rec}}}
\newcommand{\Pdet}{{P_{\rm det}}}
\newcommand{\Pobs}{{P}}
\newcommand{\Pcmb}{{P_{\rm cmb}}}
\newcommand{\Pfg}{{P_{\rm fg}}}

\newcommand{\rec}{{{\rm rec}}}
\newcommand{\Det}{{\rm{det}}}
\newcommand{\fg}{{\rm{fg}}}
\newcommand{\red}{{{\rm uncorr}}}

\newcommand{\g}{{\mathcal{G}}} 
\newcommand{\pr}{{\mathcal{P}}}
\newcommand{\mask}{{W}}
\newcommand{\wien}{{D_p}}

\title[The axis of evil - a polarization perspective] 
{The axis of evil - a polarization perspective} 
\author[M. Frommert \& T.~A. En{\ss}lin]
{M. Frommert \& T.~A. En{\ss}lin\\
Max-Planck-Institut f\"ur Astrophysik,
Karl-Schwarzschild-Stra{\ss}e 1, D-85748 Garching bei M\"unchen, Germany\\
mona@mpa-garching.mpg.de\\
}

\begin{document}

\date{Accepted ??? Received ???; in
  original form ???}
\pagerange{\pageref{firstpage}--\pageref{lastpage}} \pubyear{2008}
\maketitle
\label{firstpage}

\begin{abstract}

We search for an unusual alignment of the preferred axes of the quadrupole
and octopole, the so-called {\it axis of evil}, in the CMB temperature
and polarization data from WMAP. We use the part of the
polarization map which is uncorrelated with the temperature map
as a statistically independent probe of the axis of evil,
which helps to assess whether the latter has a cosmological
origin or if is a mere chance fluctuation in the temperature.
Note, though, that for certain models creating a preferred axis
in the temperature map, we would not expect to see the axis in the
uncorrelated polarization map.
We find that the axis of the quadrupole of the uncorrelated
polarization map roughly aligns with the axis 
of evil within our measurement precision, whereas the axis of the
octopole does not. However, with our measurement uncertainty, the
probability of such a scenario to happen by chance in an isotropic
universe is of the order of 50 per cent.
We also find that the so-called cold spot present
in the CMB temperature map is even colder in the 
part of the temperature map which is uncorrelated with the polarization,
although there is still a large uncertainty in the latter.
Therefore, our analysis of the axis of evil and a future analysis of the
cold spot in the uncorrelated temperature data will strongly benefit
from the polarization data expected from the {\it Planck} satellite. 

\end{abstract}



\section{Introduction} \label{intro}

A major assumption of modern day cosmology is the
cosmological principle, which states that the Universe is homogeneous and
isotropic on large scales. 
The observed isotropy of the Cosmic Microwave Background (CMB) is one of
the strongest evidences supporting the cosmological principle.

However, in recent years, there have been claims of anomalies detected
in the CMB temperature map with considerable significance, which seem
to break statistical isotropy of the temperature fluctuations and thus
to question the cosmological principle. Several groups
\citep{oliveira,abramo,land_2,kumar,schwarz} 
claim to have found a strong alignment between the preferred axes of
the quadrupole and the octopole, which is commonly referred to as the
{\it axis of evil}. Others \citep{bernui,eriksen,hoftuft} have
found a significant 
power asymmetry between the northern and southern ecliptic hemisphere,
and some weaker anomalies have been found for the low multipoles beyond the
octopole \citep{copi_2,land_1,abramo,pereira}.
However, the existence of such an isotropy breaking in the CMB
temperature map is strongly under debate, and also 
negative results have been published \citep{souradeep_1,magueijo_occam}.
The claims of the existence of a preferred direction in the CMB
temperature map have led to a discussion about whether this
is simply due to a chance fluctuation in the CMB temperature
map, if it can be blamed on local structures or on systematics in
the measurement, or whether it is actually due to a preferred
direction intrinsic to our Universe 
\citep{copi,klaus1,
  klaus2,symm_based,bayesian_groneboom,morales,vielva_anomalies,silk,gao,ackerman,copi_4,copi_3,hansen,hansen_2,prunet,tess,tess_2,bernui_2,wiaux}.

The polarization fluctuations of the CMB, just as its
temperature fluctuations, have 
their origin in the primordial gravitational potential.
The polarization should thus exhibit similar 
peculiarities as the temperature, provided they 
are due to some preferred direction 
intrinsic to the geometry of the primordial Universe.
Note that this is not generic
to every theoretical model creating anomalies in the temperature map.
For example, if the peculiarities in the temperature maps are due to a
secondary effect on the CMB such as the integrated Sachs-Wolfe
effect, we would not expect them to be present in the polarization maps
\citep{hu_polarization}. 
The search for anomalies in the CMB polarization map is still in
its initial stage, due to the high noise-level in the available full-sky
polarization map from the {\it Wilkinson Microwave Anisotropy Probe}
(WMAP). \cite{souradeep_1} have found some evidence for
anisotropies in the WMAP polarization data using the method proposed in
\cite{basak}. However, they state that the anisotropies are likely due
to observational artifacts such as foreground residuals, and that
further work is required in order to confirm a possible cosmic origin.

Given that the polarization map is correlated with the temperature map,
it is not a statistically independent probe of the anomalies which have
been found in the temperature map. If the observed anomalies
were due to a chance fluctuation in the temperature map, this chance
fluctuation could also be present in the polarization maps, due to the
correlation between the two \citep{hu_polarization}.
In this work, we split the WMAP polarization map into a part
correlated with the temperature map, $\Pknown$, and a part
uncorrelated with the latter, $\Pred$. 
We obtain the part of the polarization map which is
correlated with the temperature map by simply
translating the temperature map into a polarization map, using their
cross-correlation.
The part of the polarization map which is uncorrelated with the
temperature map serves as a statistically independent probe of the
above-mentioned anomalies. Chance fluctuations in the
temperature maps do not affect the uncorrelated
polarization map, so that a detection of the anomalies
in the latter would be a hint to an actual cosmological origin of them.
Note, though, that this does not have the power to exclude
residual foregrounds or systematics as potential origins for the
anomalies.

Similarly, we split the WMAP temperature map
into a part correlated with the polarization map, $\Tknown$, and an
uncorrelated map, $\Tred$. 
If the anomalies detected in the CMB 
temperature map are of genuine cosmological origin, they should be
present in the correlated and the uncorrelated parts of both the
temperature and polarization map. For convenience, the four resulting maps are
summarised and briefly described in Table \ref{tab:axes}.
\begin{table}
\centering
\begin{tabular}{l|l|l|l|l|l}
map & explanation & eq. & multipole & $(l,b)$ & $\sigma$  \\
\hline
$\Pknown$ & "$T \rightarrow P$" & (\ref{def_Pknown}) & quadr &
$(-117^\circ,60^\circ)$ & - \\ 
        &&& oct & $(-124^\circ,66^\circ)$ & - \\
$\Pred$ & "$\Pobs-\Pknown$" & (\ref{wiener_red}) & quadr &
$(-79^\circ,36^\circ)$ & $42^\circ$  \\ 
        &&& oct & $(-17^\circ,0^\circ)$ & $48^\circ$  \\ 
$\Tknown$ & "$P \rightarrow T$" & (\ref{T_known_2}) & quadr &
$(-73^\circ,42^\circ)$ & 
$42^\circ$  \\  
        &&& oct & $(-17^\circ,-19^\circ)$ & $37^\circ$ \\ 
$\Tred$ & "$\Tcmb - \Tknown$" & (\ref{def_Tred}) & quadr &
$(-107^\circ,42^\circ)$ & $33^\circ$  \\  
        &&& oct & $(-112^\circ,54^\circ)$ & $10^\circ$  \\ 
\end{tabular}
\caption{Axes and their uncertainties for the four different maps in Galactic
  coordinates. The large
  errors are due to the effects of the mask, residual foregrounds and 
  the detector noise in the WMAP polarization data.}
\label{tab:axes}
\end{table} 

In this work, we focus on using the uncorrelated polarization map to
probe the axis of evil. 
In order to define the
preferred axis of the multipoles, we use a statistic proposed by
\cite{oliveira}, which is the axis around which the angular momentum
dispersion is maximised for a given multipole {\it l}. We note that we
have to mask out about 25 per cent of the sky in the WMAP polarization
data in order to reduce Galactic foregrounds. Furthermore, 
the polarization data are highly contaminated by detector
noise and residual foregrounds even outside the mask.
We therefore perform a Wiener filtering of the polarization data
before determining the preferred axes, in order to reduce the noise
contained in the maps. However, we still expect a large uncertainty in
the axes, which we obtain by running Monte Carlo (MC) simulations
conditional on the data. The uncertainty in our axes
amounts to $\sigma \approx 45^\circ$.

We find that, for all four of the maps, the preferred axes of the
quadrupole all point in the same direction, within our
measurement precision. However, the preferred axis of the octopole of
the uncorrelated polarization map does
not align with the one of the quadrupole. The same holds for the
correlated temperature map.

In order to assess our result, we ask the following 
question. We take the axes measured in the temperature map as given,
and assume that the axes of the uncorrelated polarization map are
distributed isotropically and independently of each other. We then ask
how likely it is that at least one of these axes lies such that the
axis of the temperature map lies inside its $1\sigma$ region. This
probability amounts to about 50 per cent for currently available
polarization data. This high probability is due to the large
uncertainties we have in the axes of the uncorrelated polarization map.
The main contribution of this uncertainty comes from the
high noise-level in the polarization data rather than from the mask. We
can therefore hope that the {\it Planck} polarization data
\citep{planck} will yield much stronger constraints on the axes than
the WMAP data. 

Note that our approach to probing the axis of evil in
polarization is phenomenological, since not all theoretical models
of the primordial Universe
exhibiting anomalies in the CMB temperature map show the same
behaviour in the uncorrelated polarization map. We outline a more
thorough analysis, taking into account the predictions of the
specific models for the uncorrelated polarization map, in the
conclusions of this work.

The article is organised as follows. In section \ref{sec:wiener_pol}, we
briefly review the Wiener filter. In sections \ref{sec:split_t} and
\ref{sec:split_p}, we explain 
in detail the splitting of the WMAP temperature and polarization maps,
respectively. Section \ref{sec:axis} is devoted to determining the preferred
axes for the quadrupole and octopole for our four maps. We conclude in section
\ref{sec:conclusions}.

\section{Wiener filtering}\label{sec:wiener_pol}

\begin{figure}
 \centering
 \includegraphics[scale=0.4]{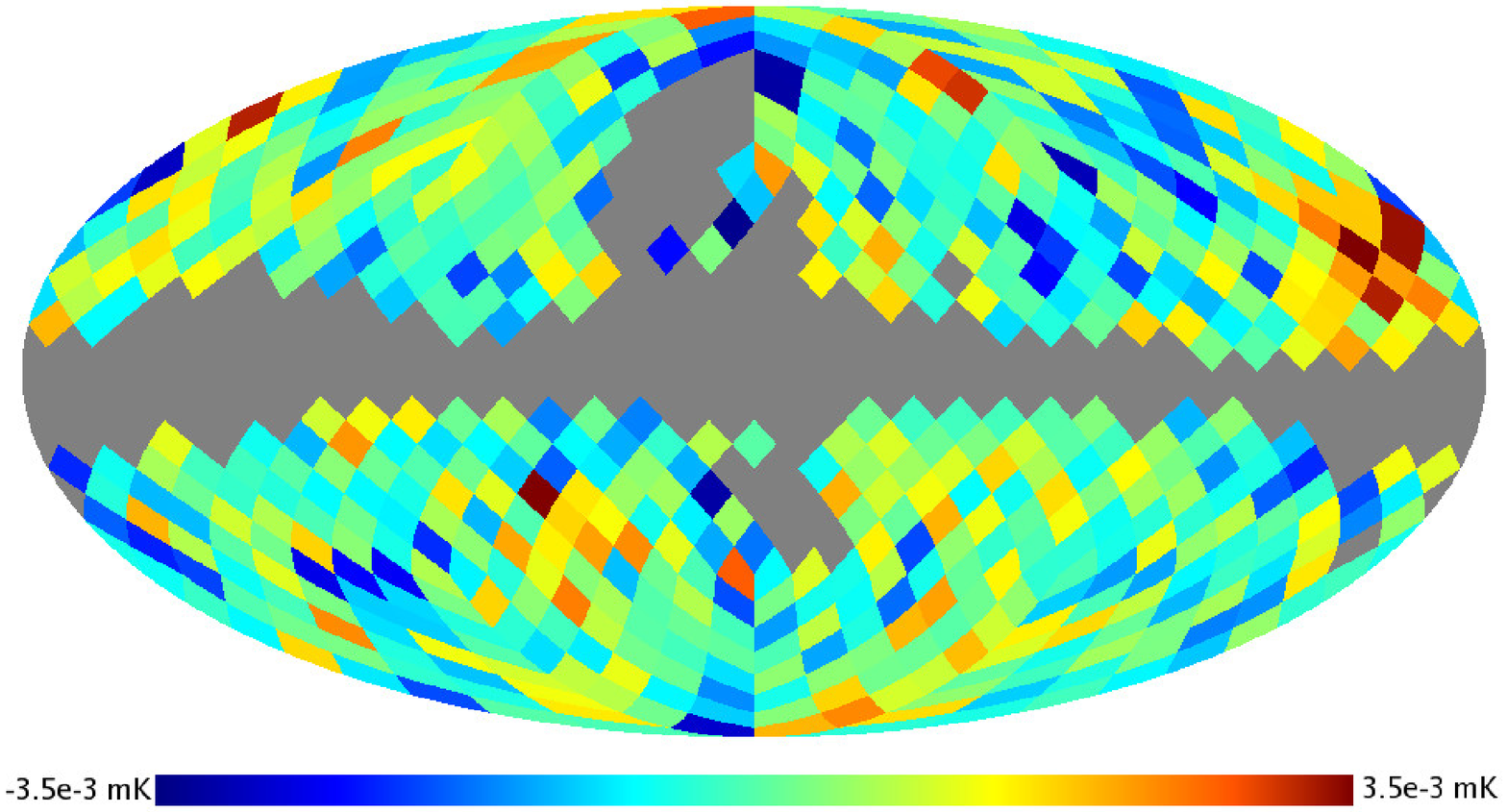}
 \includegraphics[scale=0.4]{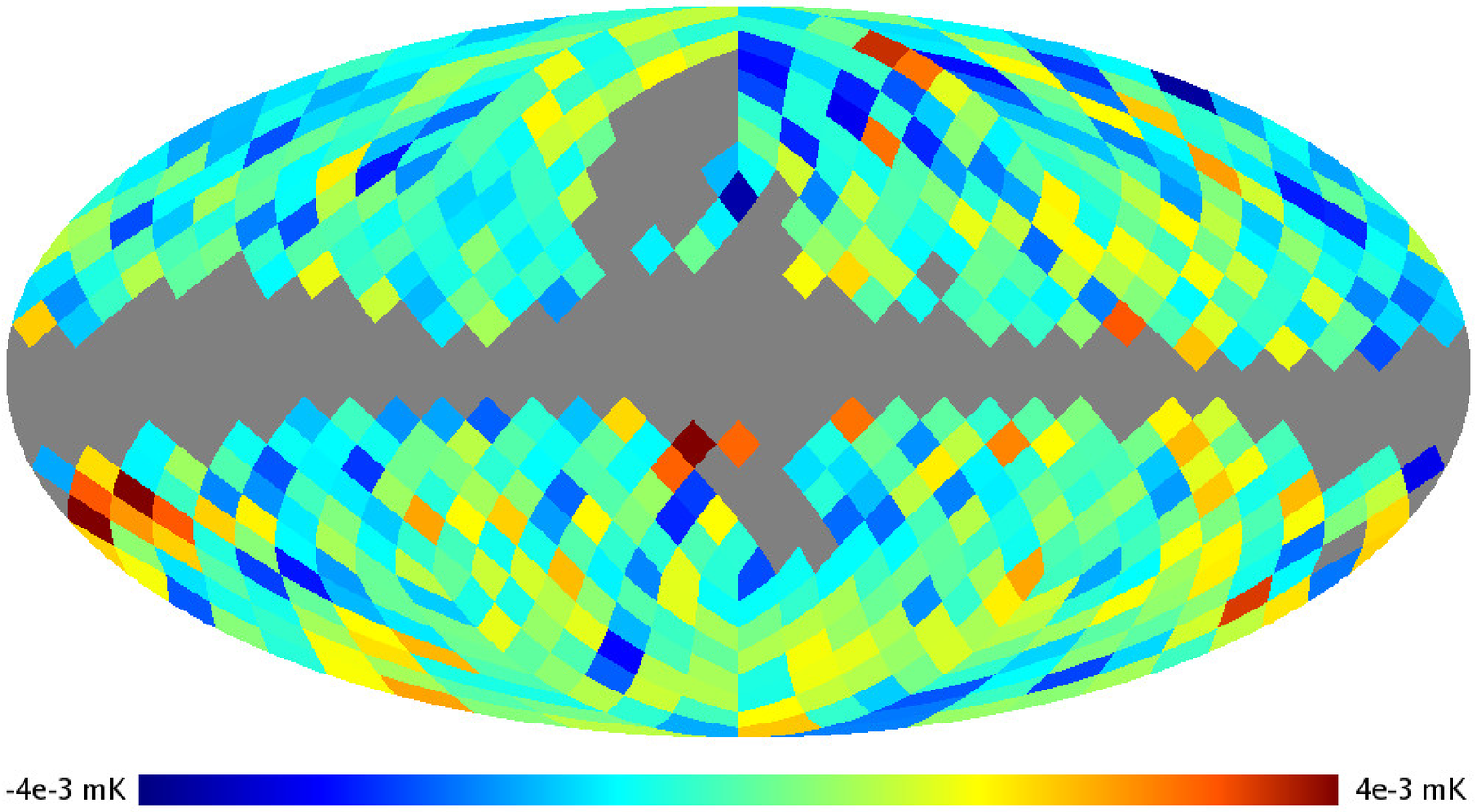}
 \caption{Observed polarization maps (linear combination of Ka, Q, and
   V band maps). Stokes Q map (top panel) and Stokes U map (bottom panel).}
 \label{fig:Q_U_maps}
\end{figure}

As we have already mentioned in the introduction, the WMAP
polarization data are highly contaminated by detector noise and
Galactic foregrounds.
The observed polarization map we use is
the linear combination of the maps of the Ka, Q, and V
frequency bands (corresponding to 33, 41, and 61 GHz), which is
used for determining the low-l polarization likelihood in the 5 year WMAP
likelihood code \citep{wmap5_hinshaw}. 
By using the linear combination of the maps, we combine the information
from different frequency bands rather than using only the information
contained in a particular band. 
Therefore, the linear combination is less contaminated by noise than the
original maps per frequency band.
We use the P06 mask \citep{page_pol} to mask out the Galactic plane in
the polarization map. 
The linear combination maps for the Stokes Q and U parameters are
shown in Fig. \ref{fig:Q_U_maps} in Galactic coordinates. 

In order to reduce the noise level, 
we perform a Wiener filtering of the observed polarization map before
translating it into the part of the temperature map which is
correlated with the polarization data. Similarly, we will perform a
Wiener filtering of the part of the polarization map which is
uncorrelated with the temperature map, as we will describe in detail later
on. 

The Wiener filter can be derived for the general data model
\be
d = Rs+n\,,
\label{wiener}
\ee
where $d$ denotes the data, $s$ the (temperature or polarization)
signal, $R$ the instrument response and $n$ 
additive noise. Let us define the signal and noise covariances,
\bea 
S &\equiv& \langle s s^\dagger \rangle_{\pr(s)}
\equiv \int \mathcal{D}s \,(s s^\dagger)\, \pr(s)\,, \\
N &\equiv& \langle n n^\dagger \rangle_{\pr(n)}
\equiv \int \mathcal{D}n \,(n n^\dagger)\, \pr(n)\,,
\eea
where the dagger denotes a transposed and complex conjugated quantity,
$\pr(s)$ and $\pr(n)$ denote the probability density functions of signal
and noise, respectively, 
and the integrals have to be taken over all pixels i, \eg
\be
\mathcal{D}s \equiv \Pi_{i=1}^{N_{\rm{pix}}} d s^i\,.
\ee 
If we assume the signal prior and the noise
distribution to be Gaussian, we obtain the signal posterior
\be
\pr(s \,|\, d) = \g\left(s-s_\rec, D \right)\,.
\label{posterior}
\ee
Here, we have defined 
\be 
\g(\chi,C) \equiv \frac{1}{\sqrt{|2\pi C|}}
\exp \left(-\frac{1}{2}\chi^\dagger\,C^{-1}\chi \right)
\ee
to denote the probability density function 
of a Gaussian distributed vector $\chi$ with zero mean, given
the cosmological parameters $p$ and the covariance matrix $C \equiv \langle
\chi \chi^\dagger \rangle$, where the averages are taken over the
Gaussian distribution $\g(\chi ,C)$. 
In eq. (\ref{posterior}), we have used the definitions
\be
s_\rec \equiv (S^{-1} + R^\dagger N^{-1} R)^{-1} R^\dagger N^{-1} d\,,
\ee
which is called the Wiener reconstruction of the signal, and
\be
D \equiv (S^{-1} + R^\dagger N^{-1} R)^{-1} \,,
\ee
which denotes the Wiener variance with which the real signal, $s$, fluctuates
around the reconstruction, $s_\rec$. A detailed derivation of the posterior
distribution, eq. (\ref{posterior}), can for example be found in
\cite{ift} or in \cite{mona}.

\section{Splitting of the temperature map}\label{sec:split_t}

In this section, we split the WMAP temperature map 
into a part correlated with the WMAP polarization map, $\Tknown$, and
a part which is not, $\Tred$. This is the same splitting which has
been done in \cite{mona_pol} in order to reduce the noise in ISW
measurements. We translate the
polarization map into the correlated part of the temperature map,
using the cross-correlation between the two. However, as we already
mentioned in the last section, before doing so we perform a Wiener
filtering of the observed polarization map in order to reduce the noise.

Our data model
for the observed polarization map $\Pobs$, which contains the Stokes Q
and U maps shown in Fig. \ref{fig:Q_U_maps}, is
\be
\Pobs
\equiv \left(
\begin{array}{c}
Q \\ U
\end{array}
\right)
\equiv \mask \left(\Pcmb + \Pdet + \Pfg \right)\,. 
\label{P}
\ee
Here, $\Pcmb$ is the intrinsic CMB polarization, $\Pdet$ and
$\Pfg$ denote the detector noise and residual foregrounds, respectively, and
we have introduced the window $\mask$ in order to describe the
mask. 

Let us define the signal covariance matrix of the CMB polarization
given the cosmological parameters $p$,
\be
S_P \equiv \langle \Pcmb \Pcmb^\dagger \rangle_{\pr(\Pcmb \,|\, p)} \,,
\label{sig_cov}
\ee
and the noise covariance matrices for the detector noise
and the residual foregrounds:
\bea \nonumber
N_\Det &\equiv& \langle \Pdet \Pdet^\dagger \rangle_{\pr(\Pdet)}\,, \\
N_\fg &\equiv& \langle \Pfg \Pfg^\dagger \rangle_{\pr(\Pfg)}\,.
\eea
The signal power spectrum (and thus $S_P$) has been computed using
{\small CMBEASY} 
\citep{cmbeasy} for the Maximum Likelihood cosmological model from
\cite{wmap5_dunkley}: \{$\Omega_b h^2 = 0.0227, \Omega_\Lambda = 0.751,
h = 0.724, \tau = 0.089, n_s = 0.961, \sigma_8 = 0.787$\}.

\begin{figure}
 \centering
 \includegraphics[scale=0.4]{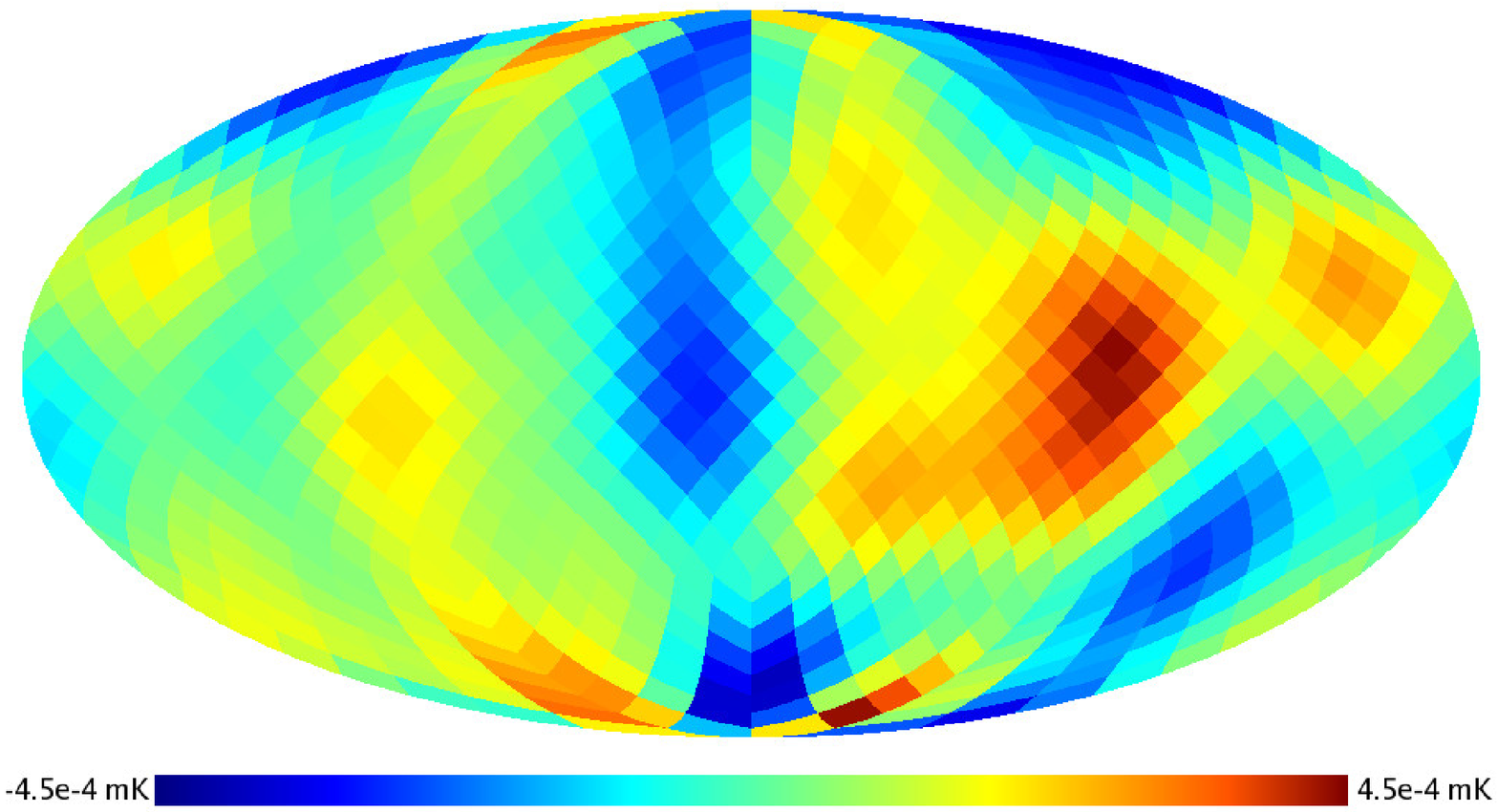}
 \includegraphics[scale=0.4]{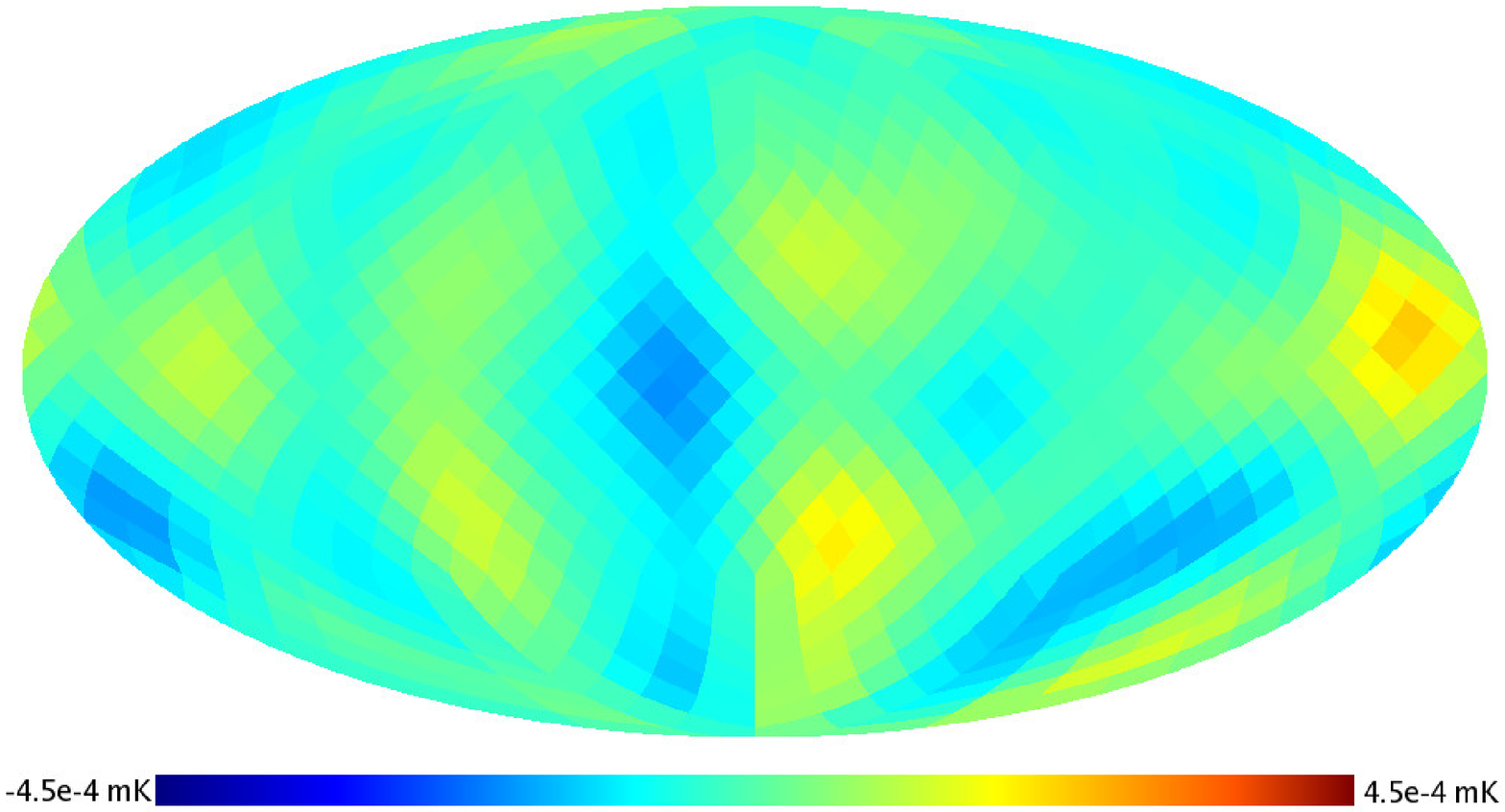}
 \includegraphics[scale=0.4]{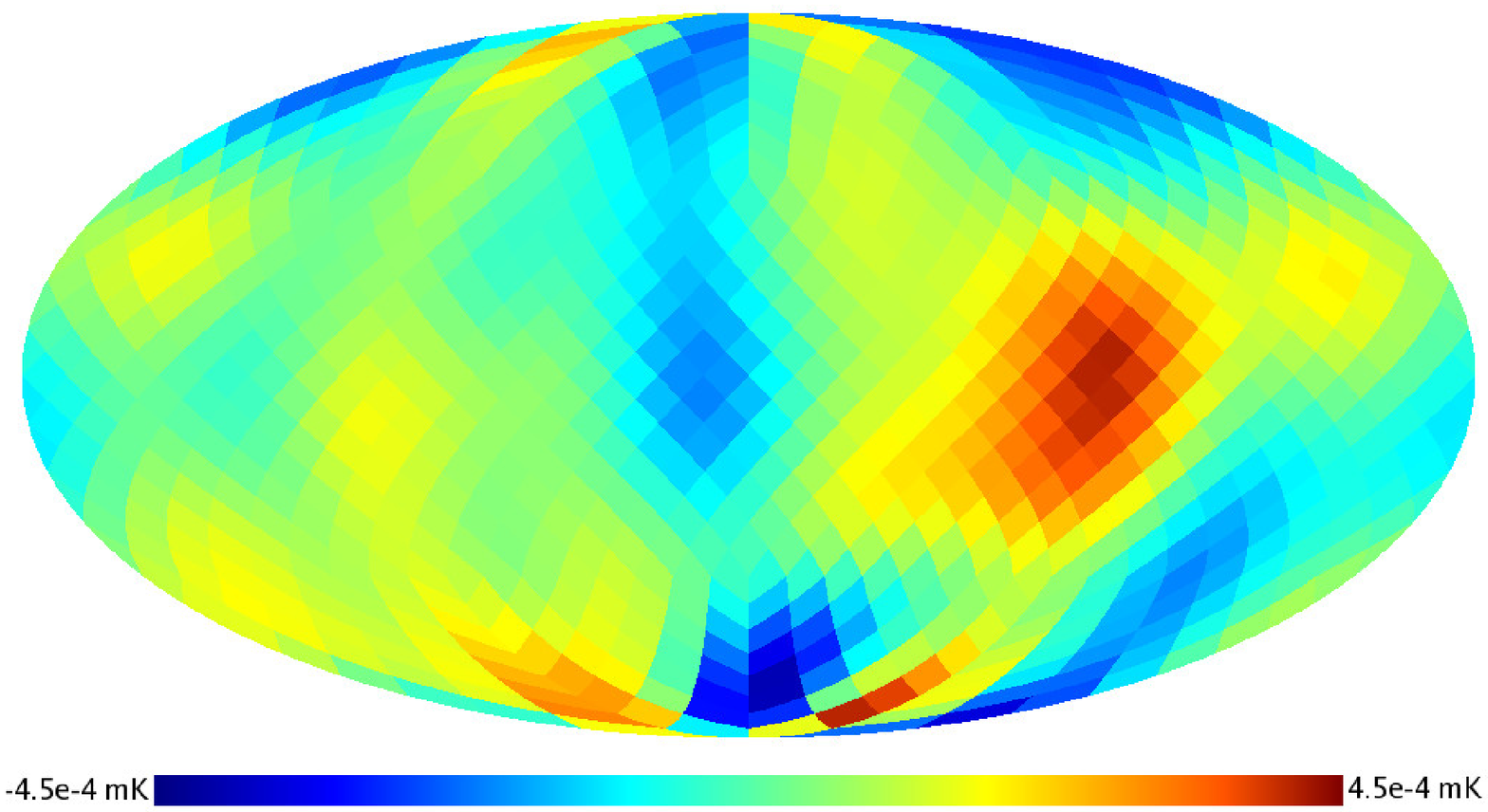}
 \caption{Stokes Q part of the following polarization maps:
   {\bf Top panel:} Wiener filtered polarization map, $\Prec$.
   {\bf Middle panel:} Part of the polarization map correlated with the
   temperature map, $\Pknown$. {\bf Bottom panel:} Part of the polarization map
   uncorrelated with the temperature map, $\Pred$. The colour scale is
   the same in all maps. 
 } 
 \label{fig:Q_wiener}
\end{figure}

\begin{figure}
 \centering
 \includegraphics[scale=0.4]{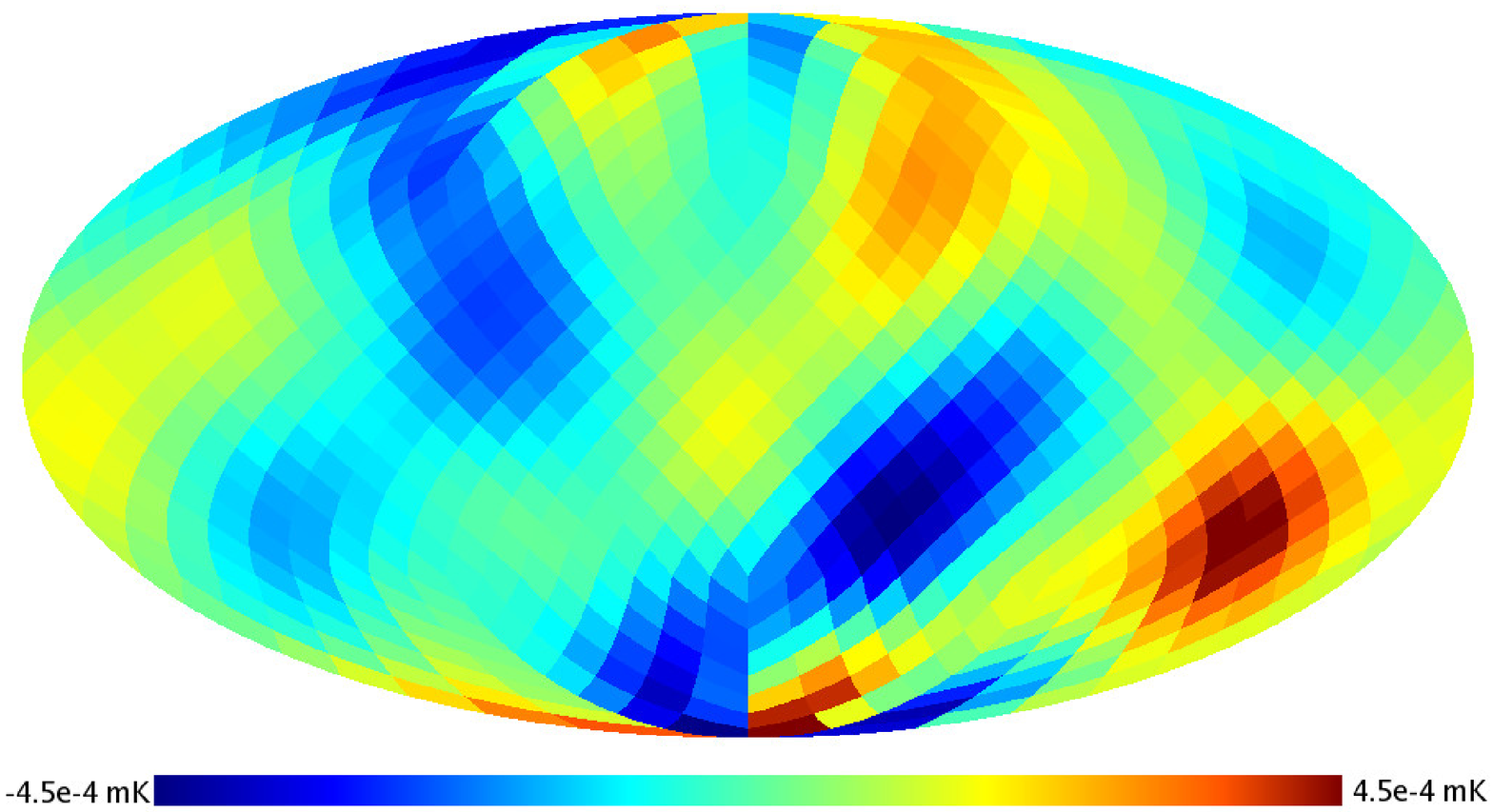}
 \includegraphics[scale=0.4]{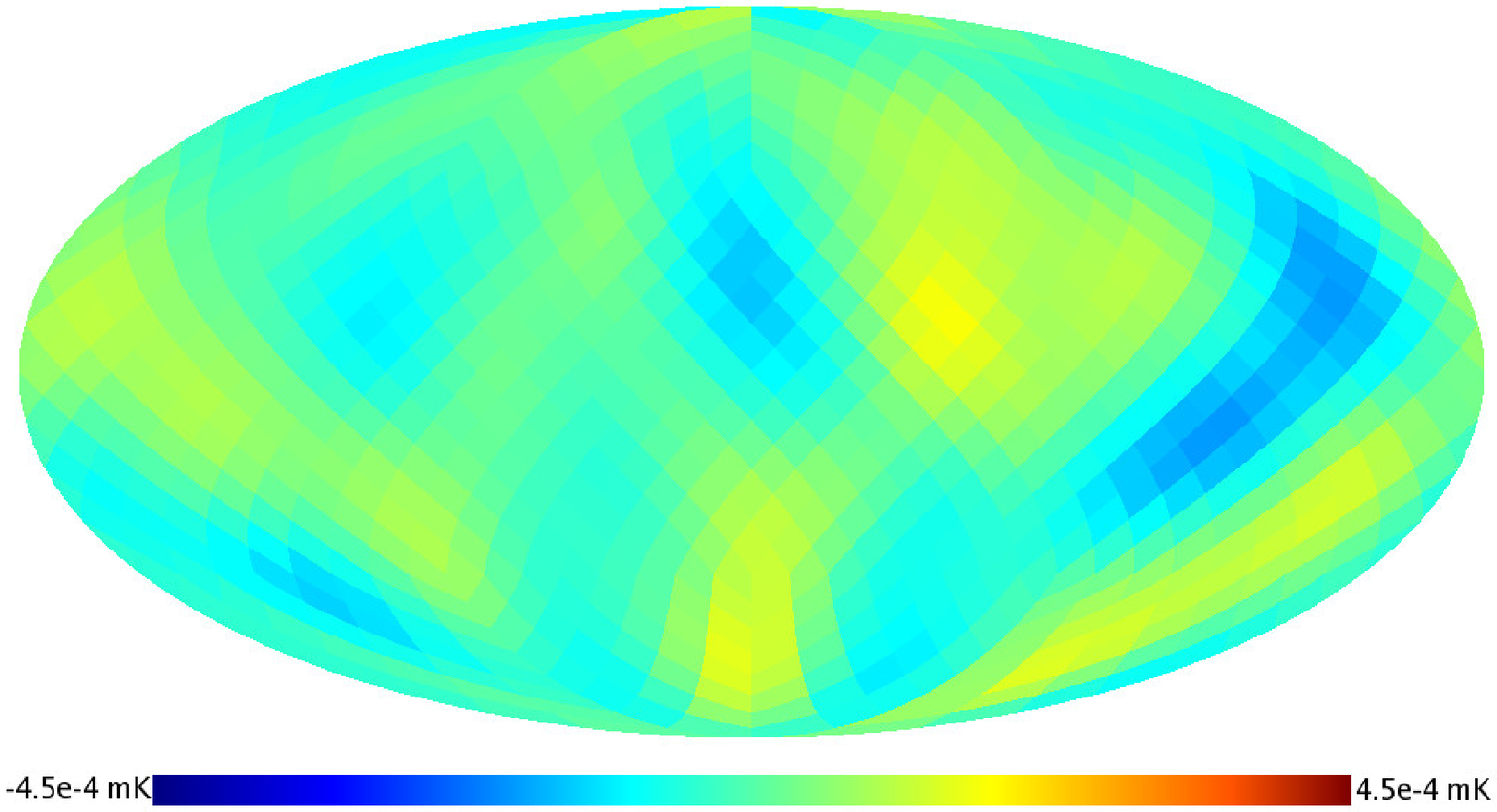}
 \includegraphics[scale=0.4]{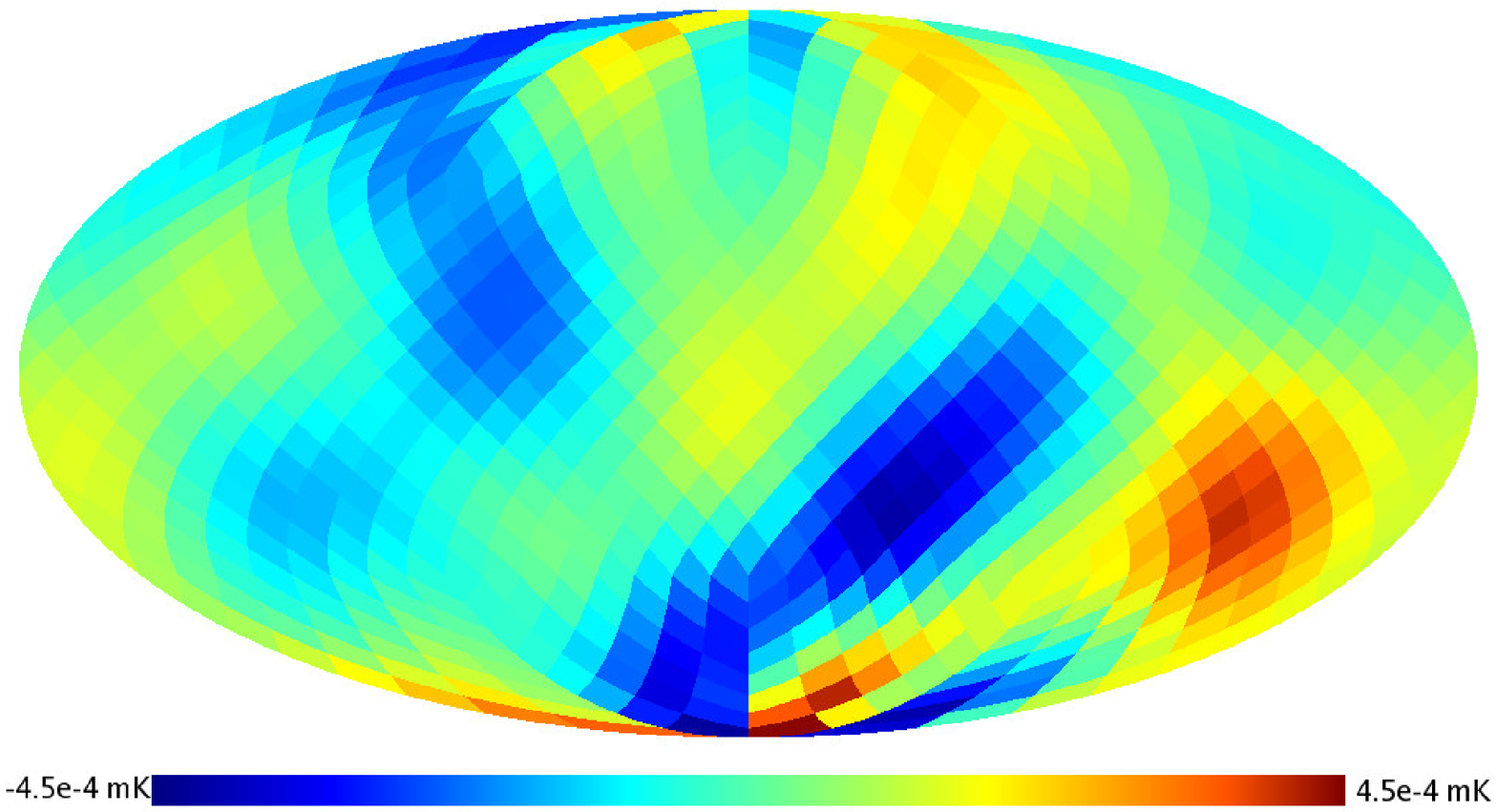}
 \caption{Stokes U part of the following polarization maps:
   {\bf Top panel:} Wiener filtered polarization map, $\Prec$.
   {\bf Middle panel:} Part of the polarization map correlated with the
   temperature map, $\Pknown$. {\bf Bottom panel:} Part of the polarization map
   uncorrelated with the temperature map, $\Pred$. The colour scale is
   the same in all maps.
 }
 \label{fig:U_wiener}
\end{figure}

In order to derive the Wiener filter for $\Pobs$, let us define the noise,
\be
n \equiv \mask \left(\Pdet + \Pfg\right)\,,
\ee
for which the noise covariance is then
\be
N_P \equiv \langle n\, n^\dagger \rangle_{\pr(n)} = \mask (N_\Det +
N_\fg) \mask^\dagger\,, 
\label{N_P}
\ee
where we have assumed that $\Pdet$ and $\Pfg$ are uncorrelated. We
take the total noise covariance, $N_P$, for the observed
polarization map from the WMAP code.
We further identify $\Pcmb$ with the signal $s$, the mask $\mask$
with the response $R$, and $\Pobs$ with the data $d$. 
With these definitions, we have translated our data model,
eq. (\ref{P}), into the one given in eq. (\ref{wiener}).
If we assume the noise $n$ and the signal $\Pcmb$ to be
Gaussian distributed\footnote{The assumption of
  Gaussianity holds well for the detector noise $\Pdet$ and the signal
  $\Pcmb$. For the residual Galactic foregrounds, this assumption is
  probably less accurate.}, we therefore obtain the posterior distribution
for the signal 
\be
\pr(\Pcmb \,|\, \Pobs, p) = \g \left(\Pcmb - \Prec, \wien \right)\,,
\label{wiener_pol}
\ee
with
\be
\Prec \equiv (S_P^{-1} + \mask^\dagger N_P^{-1} \mask)^{-1}
\mask^\dagger N_P^{-1} \Pobs\,,
\label{wiener_P}
\ee
which is the Wiener reconstruction of the polarization map, and
\be
D_p \equiv (S_P^{-1} + \mask^\dagger N_P^{-1} \mask)^{-1}\,,
\label{wiener_var}
\ee
which denotes the Wiener variance.
We show the Stokes Q and U maps of the Wiener filtered polarization map
$\Prec$ in the top panels of Fig. \ref{fig:Q_wiener} and
Fig. \ref{fig:U_wiener}, respectively. 
Note that only the low {\it l} modes survive the Wiener filtering, whereas the
higher {\it l} modes are strongly suppressed due to the high noise-level they
contain. 

We now split the WMAP temperature map into a part
correlated with the polarization map, $\Tknown$, and a part
uncorrelated with the latter, $\Pred$. We 
use the Wiener filtered polarization map $\Prec$,
which is of resolution NSIDE=8, and the internal linear
combination (ILC) temperature map \citep{wmap5_gold}, which we have
smoothed with a Gaussian beam of FWHM=$18.3^\circ$ and
downgraded to the same resolution. Among the
different WMAP temperature maps, the ILC is the one for which the 
alignment of the low multipoles is least
contaminated  by Galactic foregrounds \citep{gruppuso}.
When working on large scales, we can safely neglect the detector noise
in the temperature data \citep{afshordi_manual}.
Furthermore, we decide to neglect residual foregrounds in the temperature map.

We translate the Wiener filtered polarization map, $\Prec$,
into the correlated part of the temperature map, using the
cross-correlation between the two:
\be
\Tknown \equiv S_{T,P}\, S_P^{-1} \Prec\,,
\label{T_known_2}
\ee
where the signal covariance matrices given the cosmological
parameters, $p$, are defined as
\bea
S_{P,T} &\equiv& \langle \Pcmb \Tcmb^\dagger
\rangle_{\pr(\Tcmb,\Pcmb \,|\, p)}\,, \\
S_T &\equiv&  \langle \Tcmb \,\Tcmb^\dagger
\rangle_{\pr(\Tcmb \,|\, p)}\,.
\eea

The uncorrelated temperature map $\Tred$ is then obtained by
simply subtracting $\Tknown$ from $\Tcmb$:
\be
\Tred \equiv \Tcmb - \Tknown\,.
\label{def_Tred}
\ee
In Appendix \ref{sec:T_uncorr}, we prove that $\Tknown$ and $\Tred$
are indeed uncorrelated.

\begin{figure}
 \centering
 \includegraphics[scale=0.4]{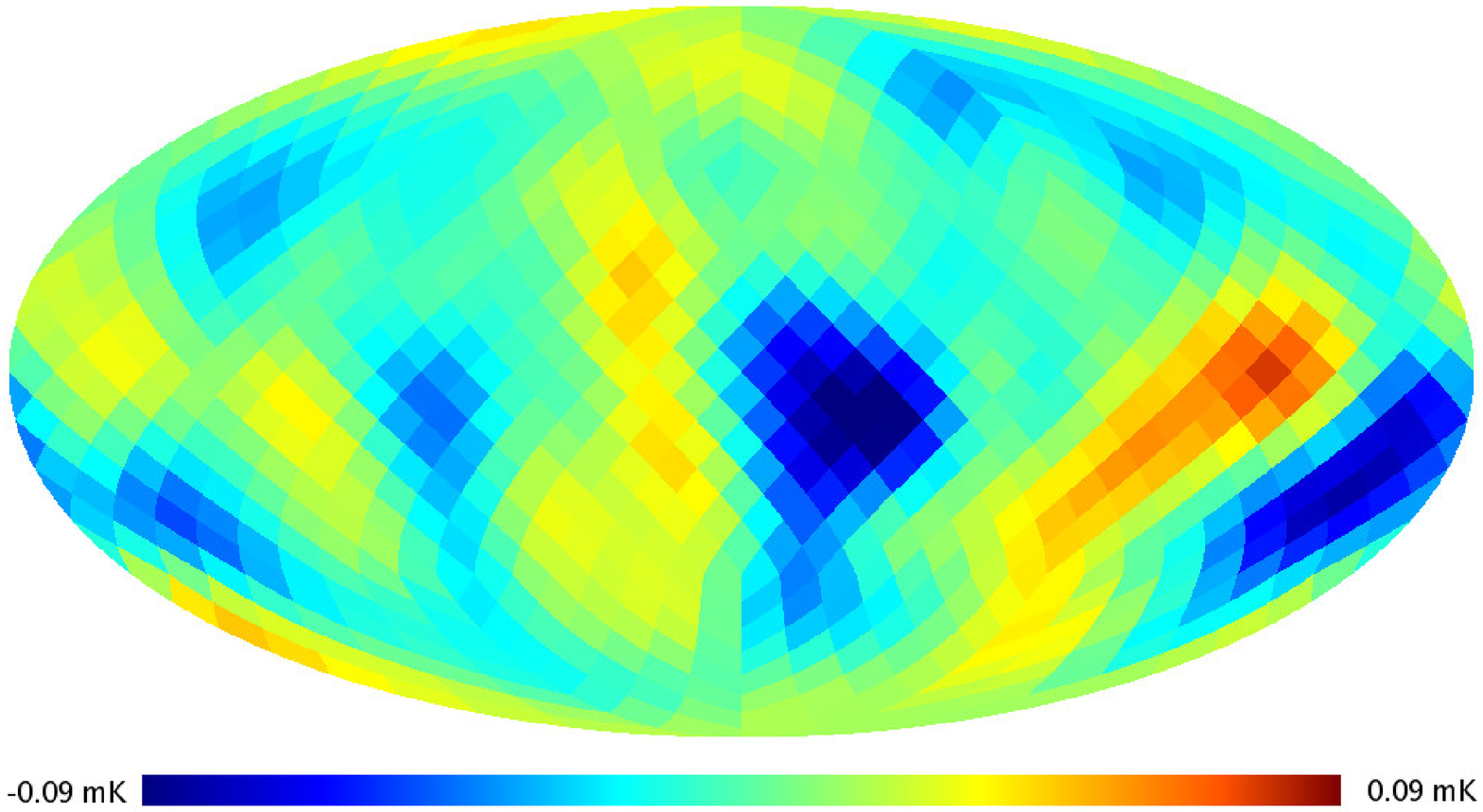}
 \includegraphics[scale=0.4]{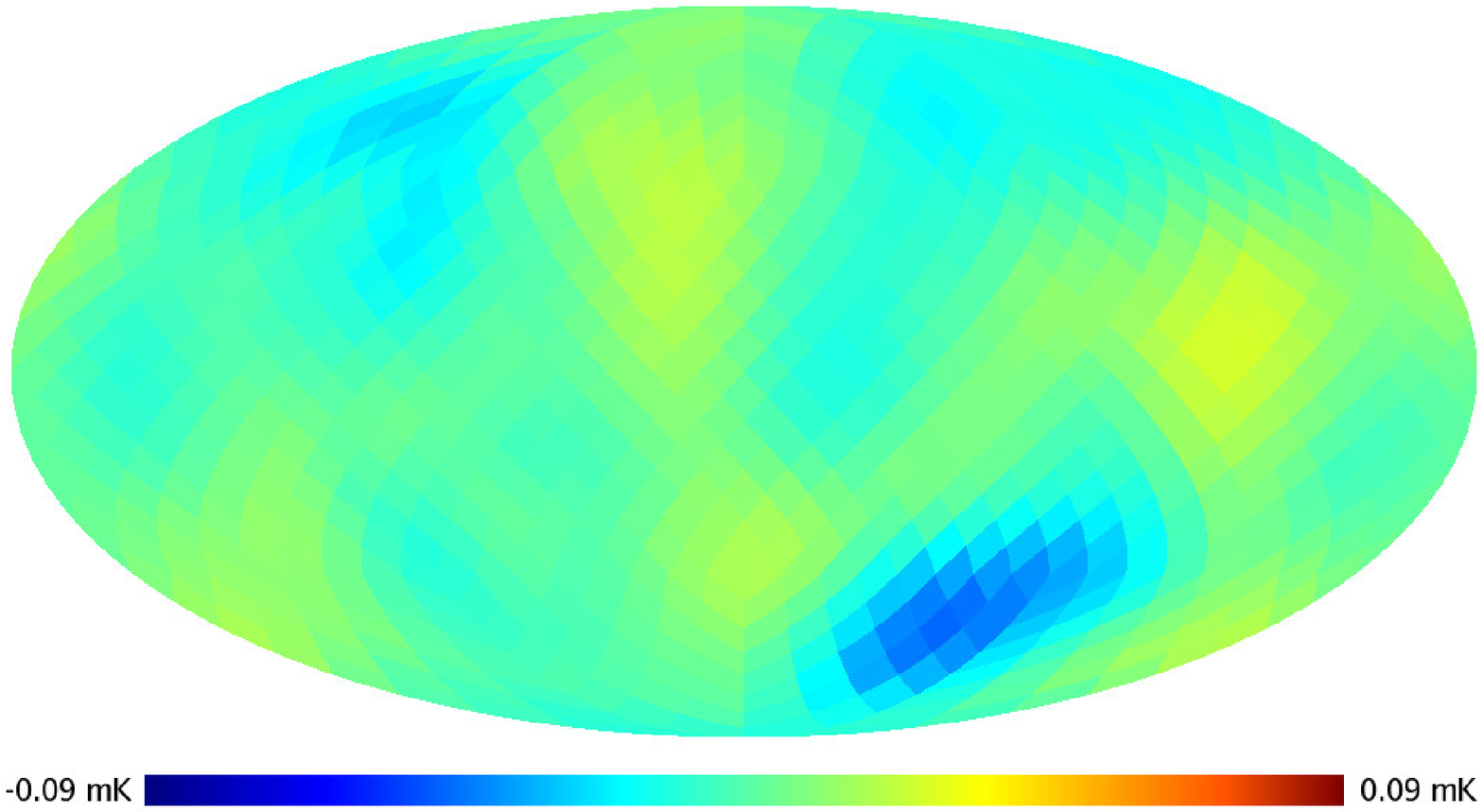}
 \includegraphics[scale=0.4]{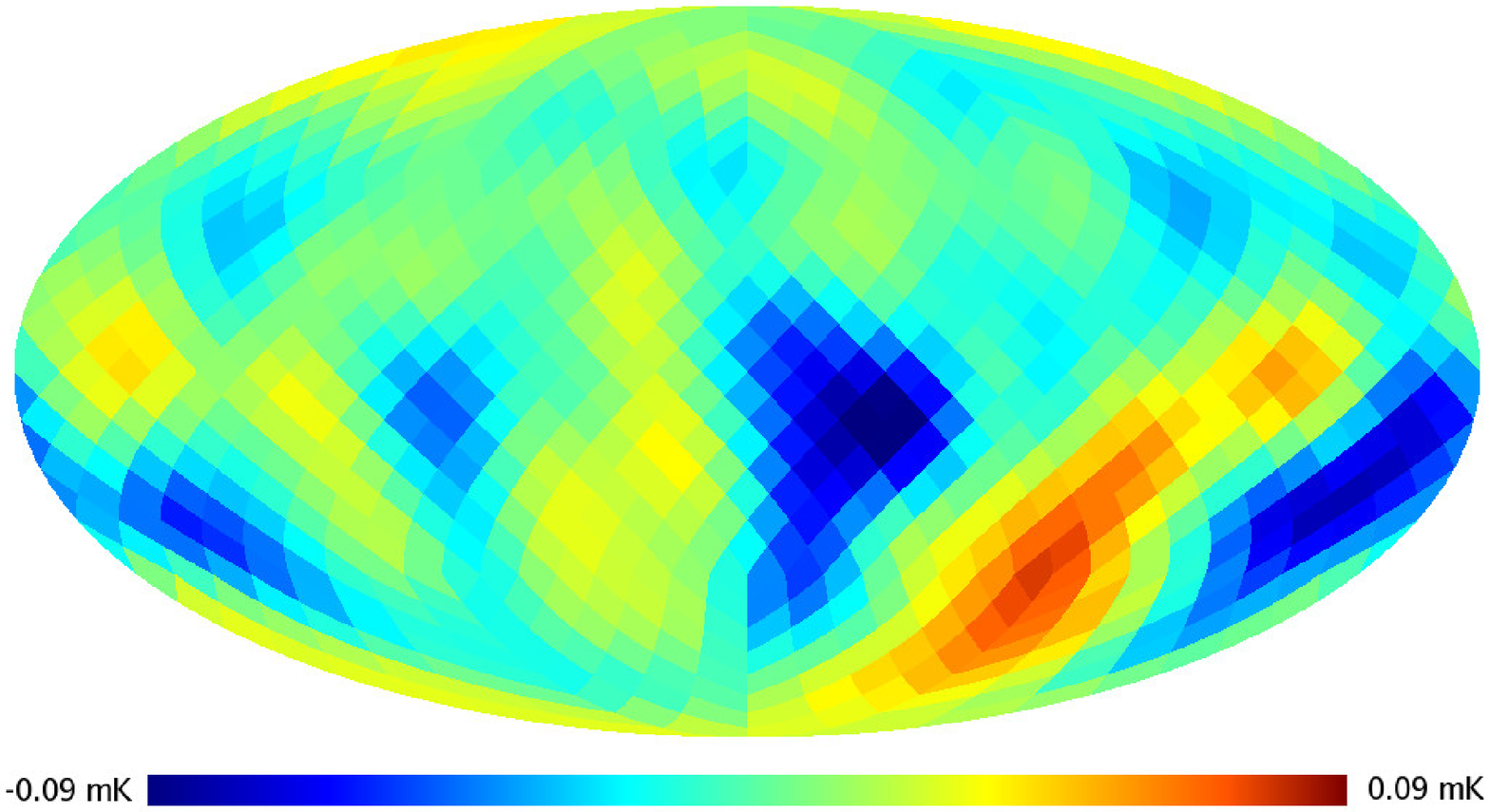}
 \caption{{\bf Top panel:} ILC map, smoothed with a beam of $18.3^\circ$
   and downgraded to a resolution of NSIDE=8. 
   {\bf Middle panel:} Part of the temperature map which is correlated
   with the polarization map, $\Tknown$. {\bf Bottom panel:}
   Part of the temperature map which is uncorrelated with the
   polarization map, $\Tred$. The colour scale is the same in all maps.
 }
 \label{fig:map_known}
\end{figure}

We plot $\Tcmb$, $\Tknown$, and $\Tred$ in the top, middle, and
bottom panel of Fig. \ref{fig:map_known}, respectively.
Let us first concentrate on $\Tknown$, and try to 
assess whether some of its structures could come from
Galactic foregrounds rather than being intrinsic CMB fluctuations. 
Note that this is just meant to be a quick glance on what we can
immediately pick out by eye.
Comparing $\Tknown$ with the overview over the Galactic foregrounds
published in \cite{wmap3_hinshaw}, Fig. 7,
makes us suspect that the warm region in the middle of the northern hemisphere
might be associated with the North Galactic Spur. A part of this region is
already masked out, but it is well possible that the mask should be bigger in
order to better mask out this foreground.
One might also think that the big red blob on the right hand side
of $\Tknown$, close to the Galactic plane, could be due to the Gum
Nebula.
However, plotting the two maps on top of each other reveals that the 
Gum Nebula lies further to the East than our red blob. Therefore we exclude
that the blob comes from that particular foreground.

\begin{figure}
 \centering
 \includegraphics[scale=0.4]{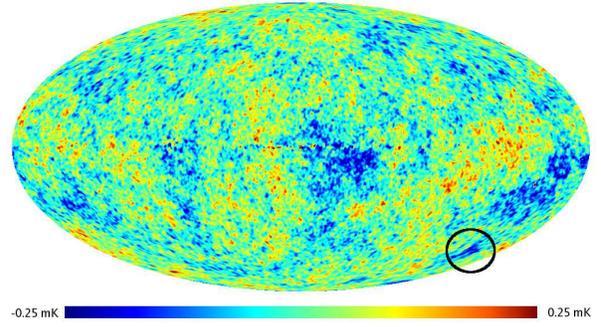}
 \caption{The {\it cold spot}, which has been found to have non-Gaussian
   characteristics, is marked in the ILC map shown here by a black circle.}
 \label{fig:ilc}
\end{figure}

Let us now compare the three maps $\Tcmb$, $\Tknown$, and $\Tred$.
In the northern Galactic hemisphere, all maps look quite
similar, apart from the hot region around the North Galactic Spur,
which is more prominent in $\Tknown$ than in the other two
maps, and which we have already commented on. However, in the western
part of the southern hemisphere, we obtain 
a strong deviation of $\Tknown$ from the ILC map. In fact, the
features in $\Tknown$ have the opposite sign to the structures in the
ILC map. This enhances the amplitudes of the features in the
western part of the southern hemisphere in $\Tred$ as compared to the
ILC map. In particular, the so-called {\it cold spot}, which has been found
to have non-Gaussian characteristics by \cite{vielva_spot}, turns out to be 
even colder in $\Tred$ than in the ILC map. The cold spot, which we
mark in the ILC map in Fig. \ref{fig:ilc} by a black circle, has
later been confirmed  to have non-Gaussian characteristics by many
others \citep[see, \eg,][]{martinez,cruz,naselsky}. It would be
interesting to redo the above-mentioned analyses of the cold spot with the
high-resolution version of $\Tred$, in order to see whether the
significance of the non-Gaussian features is even higher in that
map. A thorough analysis of the characteristics of the cold spot is
beyond the scope of this work, though, and we leave this exciting
question for future work. 
Lastly, we notice that on the large scales we are looking at, we have
much stronger deviations of the temperature towards the cold 
end of the temperature spectrum than towards the warm end, for all
three of the maps.

\section{Splitting of the polarization map}\label{sec:split_p}

We now split the WMAP polarization map into a part correlated with
the WMAP temperature map, $\Pknown$, and a part uncorrelated with
that, $\Pred$.
As before, we obtain the correlated polarization map
by simply translating the temperature map into a polarization map:
\be
\Pknown \equiv  S_{P,T} S_T^{-1} \Tcmb \,, 
\label{def_Pknown}
\ee
The Stokes Q and U maps of $\Pknown$ are shown in the middle panels of
Fig. \ref{fig:Q_wiener} and Fig. \ref{fig:U_wiener}, respectively.

In order to obtain the uncorrelated map, we would like to
subtract $\Pknown$ from $\Pcmb$: 
\be
\Predreal \equiv \Pcmb - \Pknown\,.
\ee
However, we do not know $\Pcmb$
because we only observe $\Pobs$, which is highly contaminated by
noise. Subtracting $\Pknown$ from the Wiener filtered polarization
map, $\Prec$, does not result in uncorrelated maps. We therefore
subtract $\mask \Pknown$ from the observed polarization map, $\Pobs$:
\bea \nonumber
\Predraw &\equiv& \Pobs - \mask \Pknown \\
&=& \mask \Predreal + n\,,
\label{def_Predraw}
\eea
where the noise $n$ is the same as in section \ref{sec:split_t}.
We then compute the Wiener reconstruction of the signal $\Predreal$,
with the data being $\Predraw$: 
\bea \nonumber
\Pred &=& [(S_P - S_{P,T} S_T^{-1} S_{T,P})^{-1} + \mask^\dagger N_P^{-1}
  \mask]^{-1} \\
&& \mask^\dagger N_P^{-1}\Predraw\,.
\label{wiener_red}
\eea 
Here, we have used the signal covariance
\bea \nonumber
&& \langle \Predreal\Predreal^\dagger \rangle_{\pr(\Pcmb,T
  \,|\,p)} \\ \nonumber
&=& \langle \Pcmb \Pcmb^\dagger \rangle 
- \langle \Pcmb \Tcmb^\dagger \rangle S_T^{-1} S_{T,P}  \\ \nonumber
&& - S_{P,T} S_T^{-1} \langle \Tcmb \Pcmb^\dagger \rangle 
+ S_{P,T} S_T^{-1} \langle \Tcmb \Tcmb^\dagger \rangle S_T^{-1}
S_{T,P} \\
&=& S_P - S_{P,T} S_T^{-1} S_{T,P}\,.
\eea
$\Pred$ given in eq. (\ref{wiener_red}) is
uncorrelated with $\Pknown$, as we prove in Appendix \ref{sec:P_uncorr}.
The posterior of $\Predreal$ is given by
\be
\pr(\Predreal \,|\, \Tcmb, \Pobs,p) = \g\left(\Predreal - \Pred, D_\red\right)\,,
\label{post_red}
\ee
with the Wiener variance
\be
D_\red \equiv [(S_P - S_{P,T} S_T^{-1} S_{T,P})^{-1} + \mask^\dagger N_P^{-1}
  \mask]^{-1}\,.
\label{wiener_var_red}
\ee

We show the Stokes Q and U maps of the uncorrelated polarization map,
$\Pred$, in the bottom panels of Fig. \ref{fig:Q_wiener} and
Fig. \ref{fig:U_wiener}, respectively.
Note that the symbols for the correlated and uncorrelated parts of
temperature and polarization maps are listed and briefly explained in
Table \ref{tab:axes}.

\section{The axis of evil}\label{sec:axis}

We now search for the axis of evil in the four maps $\Pknown$, $\Pred$,
$\Tknown$, and $\Tred$. Note that $\Pknown$ and $\Tknown$ have of
course the same axes as the original temperature and polarization maps,
$\Tcmb$ and $\Prec$, respectively.  
To define the preferred axis, we use a statistic proposed by
\cite{oliveira}, which has been introduced in order to quantify the
preferred direction that can be picked out in the smoothed temperature map
by eye. When 
looking at the smoothed ILC map in Fig. \ref{fig:map_known}, most of the hot
and cold blobs seem to be lying on the same plane. The quadrupole and
octopole extracted from the ILC map show the same behaviour
\citep[see, \eg,][]{oliveira}, and the planes are
roughly the same for the two multipoles. In order to quantify this alignment,
\cite{oliveira} came up with the following statistic. The temperature
maps are expanded into spherical harmonics, which are
eigenfunctions of the square and the z-component of the
angular momentum operator $\vec{L}$:
\be
\Tcmb(\vec{\hat n}) = \sum_l T_l(\vec{\hat n}) \equiv \sum_{l,m}
a_{lm}^\Tcmb Y_{lm}(\vec{\hat n})\,.
\ee
Then, for every multipole $l$, one determines the z-axis $\vec{\hat n}$
for which the expectation value of the z-component of $\vec{L}$,
$\vec{\hat n}\cdot\vec{L}$,
is maximised:
\be
\langle T_l \,|\, (\vec{\hat n} \cdot \vec{L})^2 \,|\, T_l \rangle = \sum_m
m^2 \,|\,a_{lm}^\Tcmb(\vec{\hat n})\,|\,^2\,, 
\ee
Here, $a_{lm}^\Tcmb(\vec{\hat n})$ denotes the spherical harmonic coefficient
$a_{lm}^\Tcmb$ obtained in a coordinate system with the z-axis pointing in
$\vec{\hat n}$-direction. We determine the axis $\vec{\hat n}$ by
simply rotating the z-axis into every pixel centre and checking for
the maximum, which is well feasible at our resolution. Neighbouring pixel
centres in our map differ by approximately $7^\circ$, but we will
soon see that the uncertainties in our axes are so large that it is sufficient
to check only the pixel centres as potential z-axes. We have done the same
exercise allowing the axes to point to all pixel centres of NSIDE=16
instead of NSIDE=8, and our results are robust under this change.

As we have already mentioned, the mask, residual foregrounds and
detector noise in the polarization data will result in an
uncertainty in the preferred axes. The posterior distribution of the real CMB
polarization map, $\Pcmb$, given the one we observe, $\Pobs$, is given by
eq. (\ref{wiener_pol}). $\Pcmb$ fluctuates around our Wiener
reconstruction, $\Prec$, with the Wiener variance $D_P$.

In order to obtain the uncertainties in the axes of $\Tknown$ and
$\Tred$, we have run Monte Carlo (MC) simulations, drawing
realisations of $\Pcmb$ from its posterior distribution. From these,
we obtain realisations of
\bea \nonumber
\Tknownreal &\equiv& S_{T,P}\,S_P^{-1}\Pcmb\,,\\
\Tredreal &\equiv& \Tcmb - \Tknownreal\,,
\eea
for which we then determine the preferred axes. 
The uncertainty in the axes of $\Pred$ is obtained similarly, using
the posterior distribution of $\Predreal$ given in eq. (\ref{post_red}).
Note that $\Tcmb$ and
thus $\Pknown$ are assumed to have no contributions from residual
foregrounds or detector noise, and thus no uncertainty in the preferred axes. 

\begin{figure}
 \centering
 \includegraphics[scale=0.4]{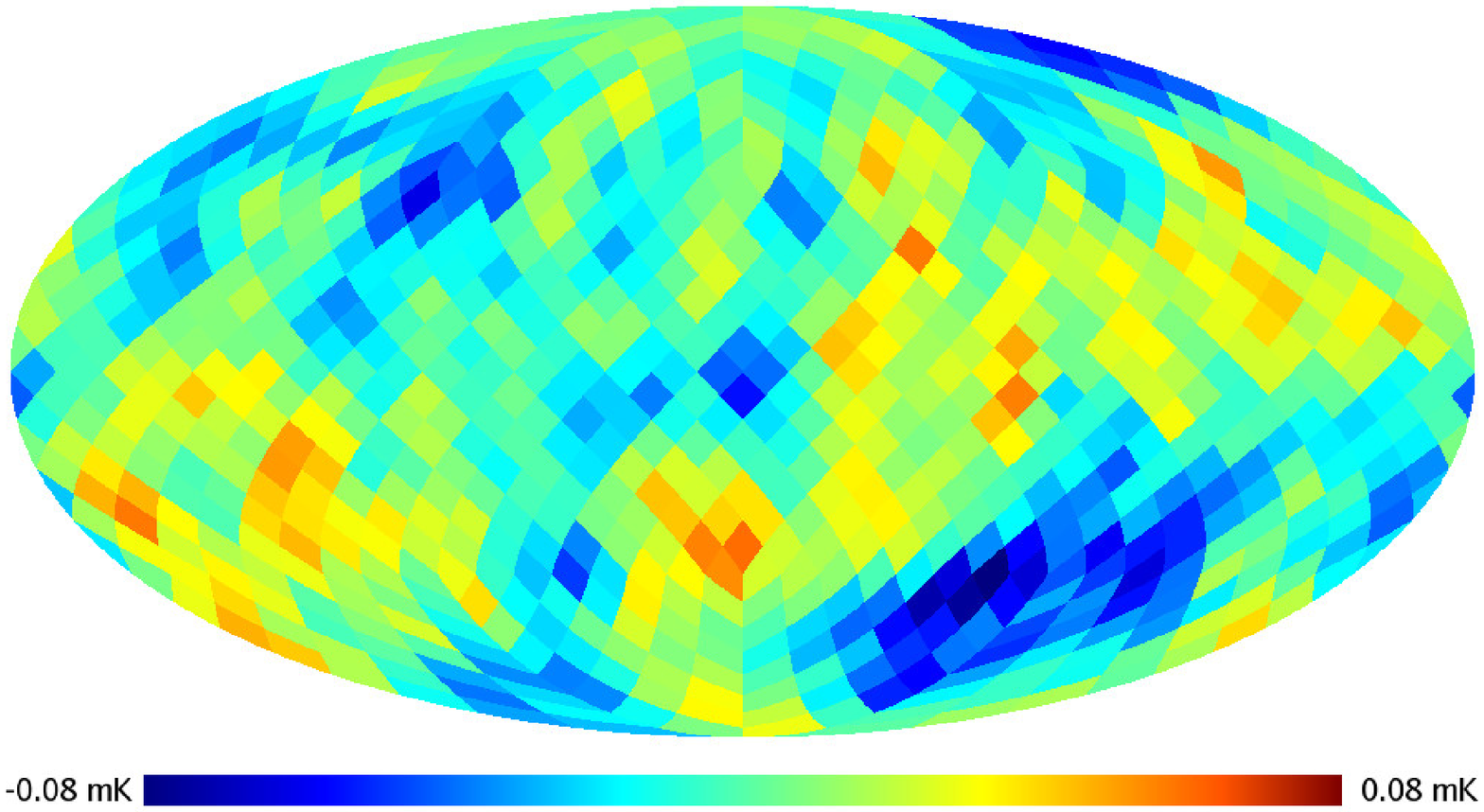}
 \caption{Wiener realisation of $\Tknownreal$}
 \label{fig:wiener_realisation}
\end{figure}

For drawing realisations from the probability
distribution in eq. (\ref{wiener_pol}), we have computed the Wiener variance
$D_P$ given in eq. (\ref{wiener_var}). We have then computed the Cholesky
factorisation $L$ of $D_P$, which is a particular form of the square-root of a
positive definite matrix:
\be
D_P = L L^\dagger.
\ee
In order to obtain our realisation, $\Pcmb$, we apply $L$ to a map $n_w$ of 
white noise, \ie a map where the temperature at every pixel is
independently drawn
from a Gaussian distribution with unit variance, and add the mean
value $\Prec$: $\Pcmb \equiv L\,n_w + \Prec$. This results in a  
map which is drawn from the distribution in eq. (\ref{wiener_pol}),
as one can easily see:
\bea \nonumber
&& \langle (\Pcmb-\Prec) (\Pcmb-\Prec)^\dagger \rangle_{\pr(n_w)} \\
&=& L \,\langle n_w
n_w^\dagger \rangle_{\pr(n_w)} \,L^\dagger = LL^\dagger = D_P.
\eea
An example of a Wiener realisation of $\Tknownreal$ in shown in
Fig. 
\ref{fig:wiener_realisation}.\footnote{We had to regularise the 
Wiener variances, eqs (\ref{wiener_var}) and (\ref{wiener_var_red}),
by adding Gaussian noise in order to make them 
positive definite. This is required by the Cholesky
factorisation. However, since the noise was added mostly on small
scales, the quadrupole and octopole remained completely unaffected by
this. In fact, our results remained unchanged under varying the
variance of the added Gaussian noise over 5 orders of
magnitude.}

\begin{figure}
 \centering
 \includegraphics[scale=0.4]{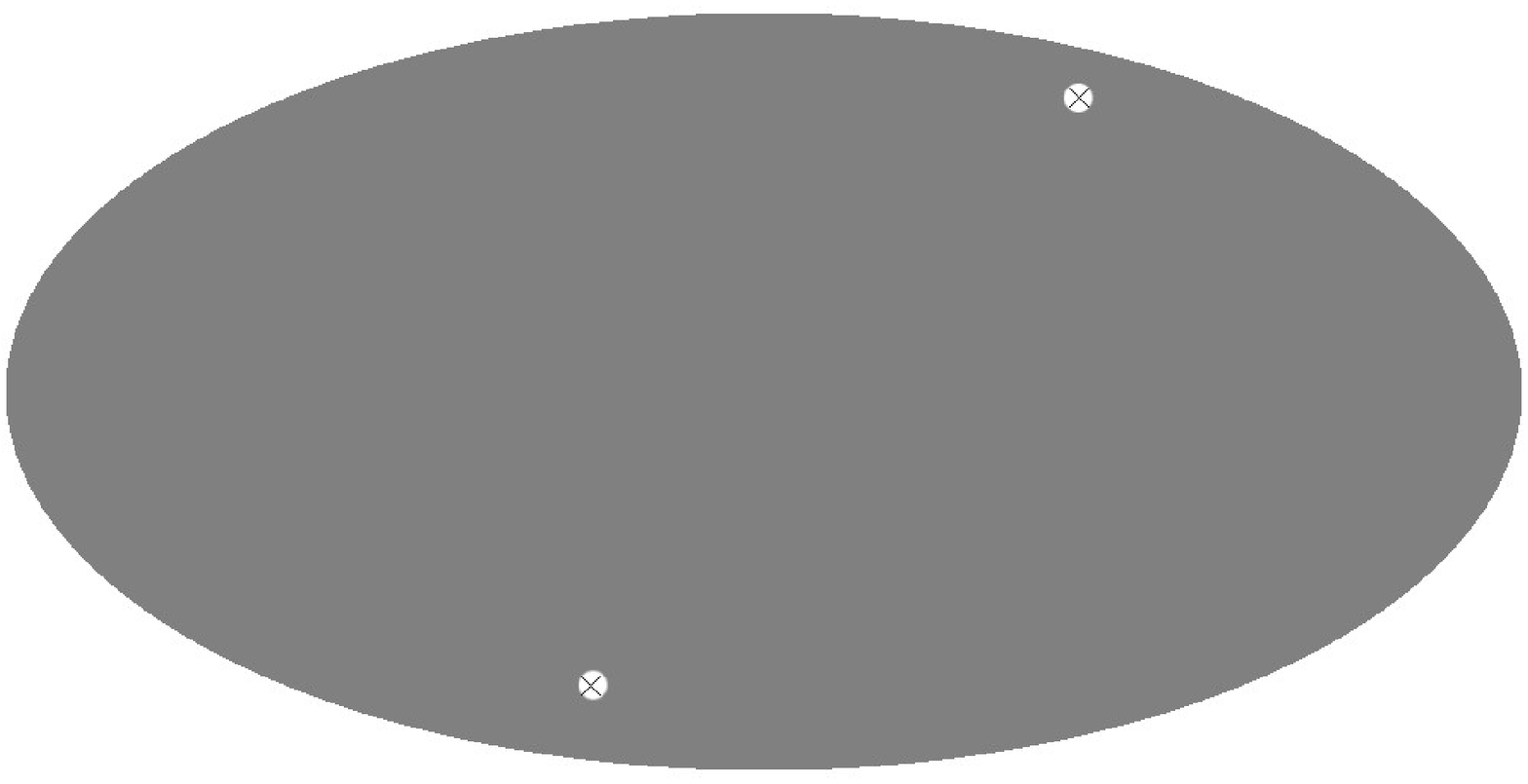}
 \includegraphics[scale=0.4]{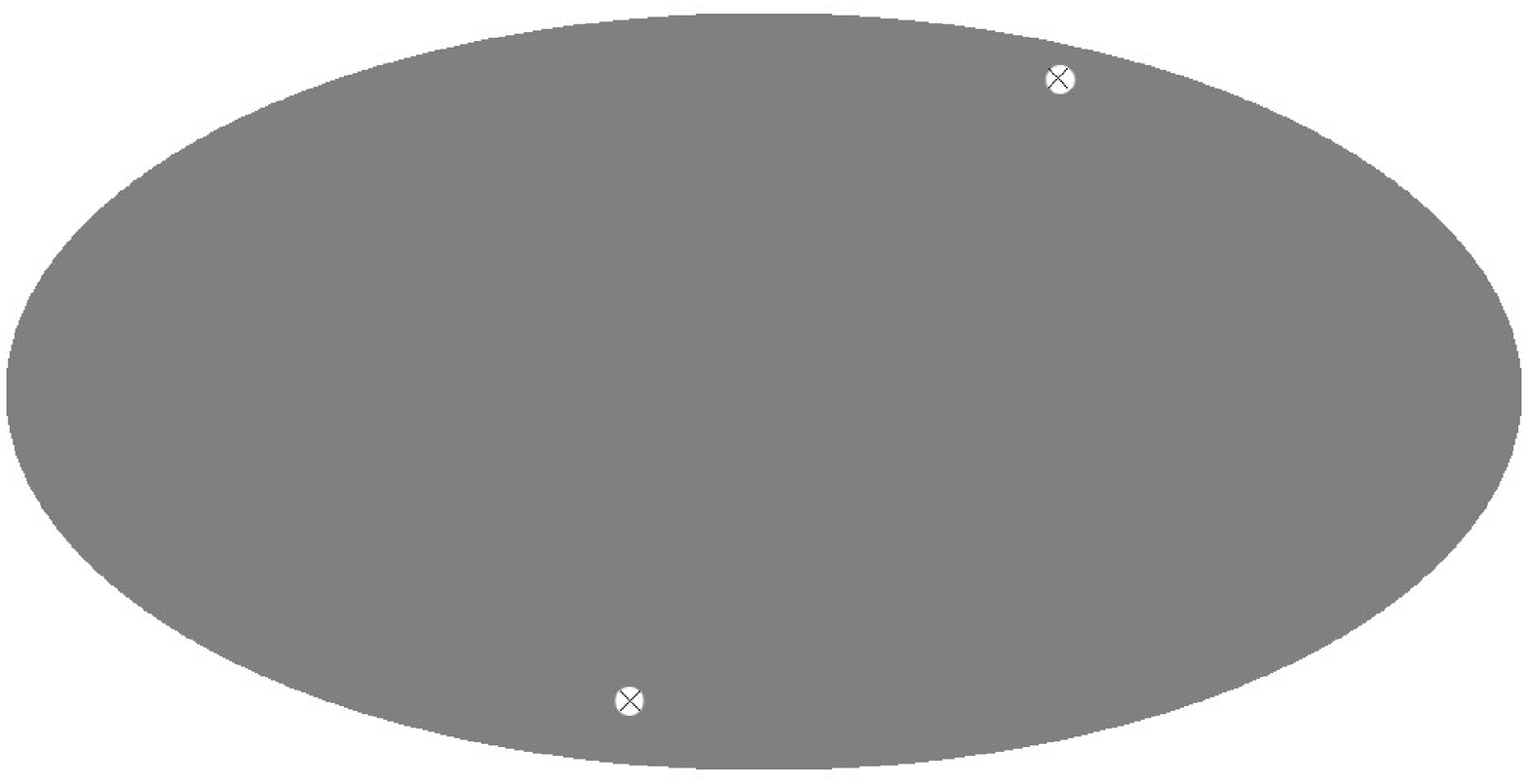}
 \caption{Preferred axis of the quadrupole (top panel) and the
   octopole (bottom panel) for $\Pknown$ and thus for the ILC map. We
   reproduce the results of \protect\cite{oliveira} within our measurement
   precision. The axes of quadrupole and octopole point in the same
   direction, which has been named the axis of evil.
 }
 \label{fig:axes_Pknown}
\end{figure}

\begin{figure}
 \centering
 \includegraphics[scale=0.4]{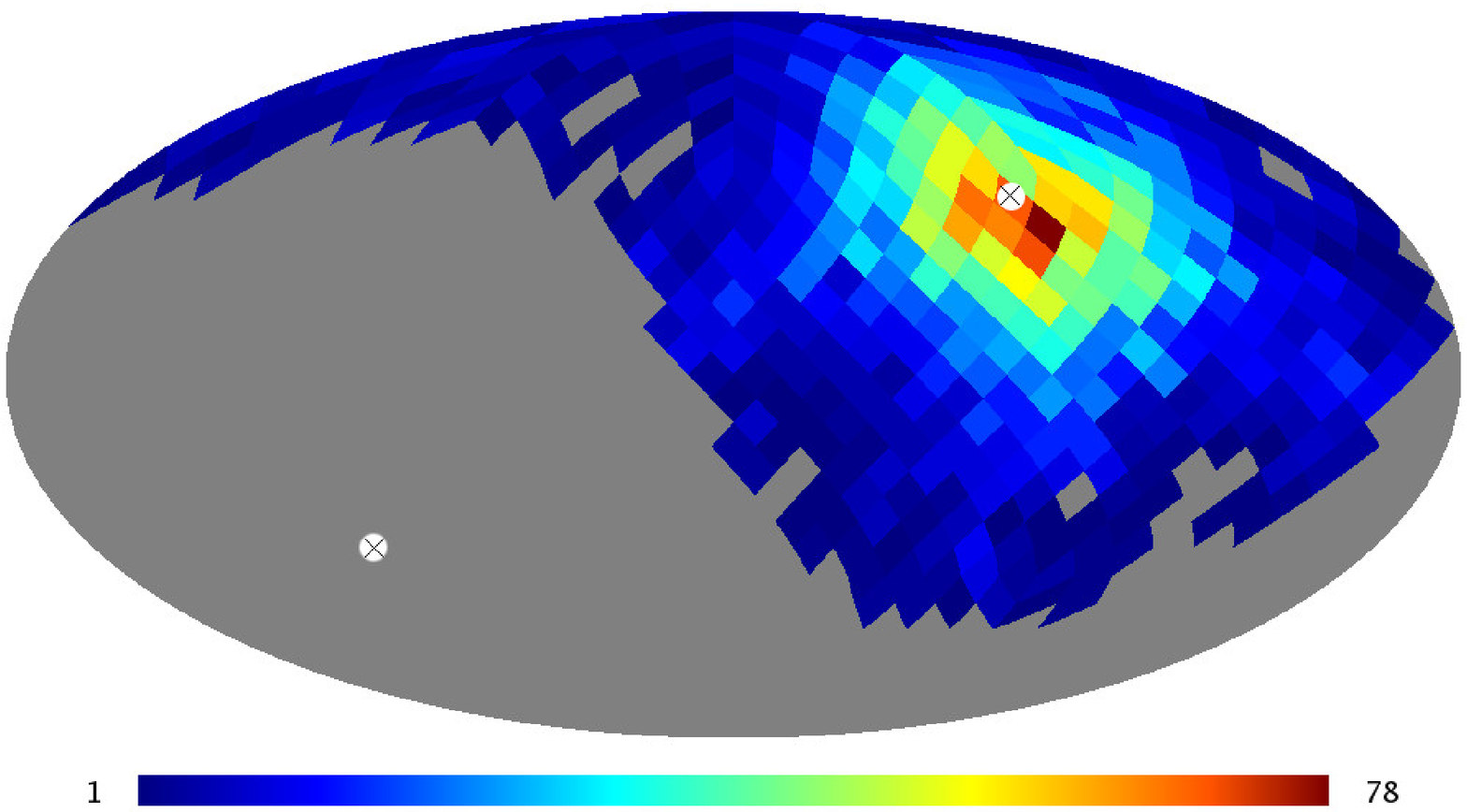}
 \includegraphics[scale=0.4]{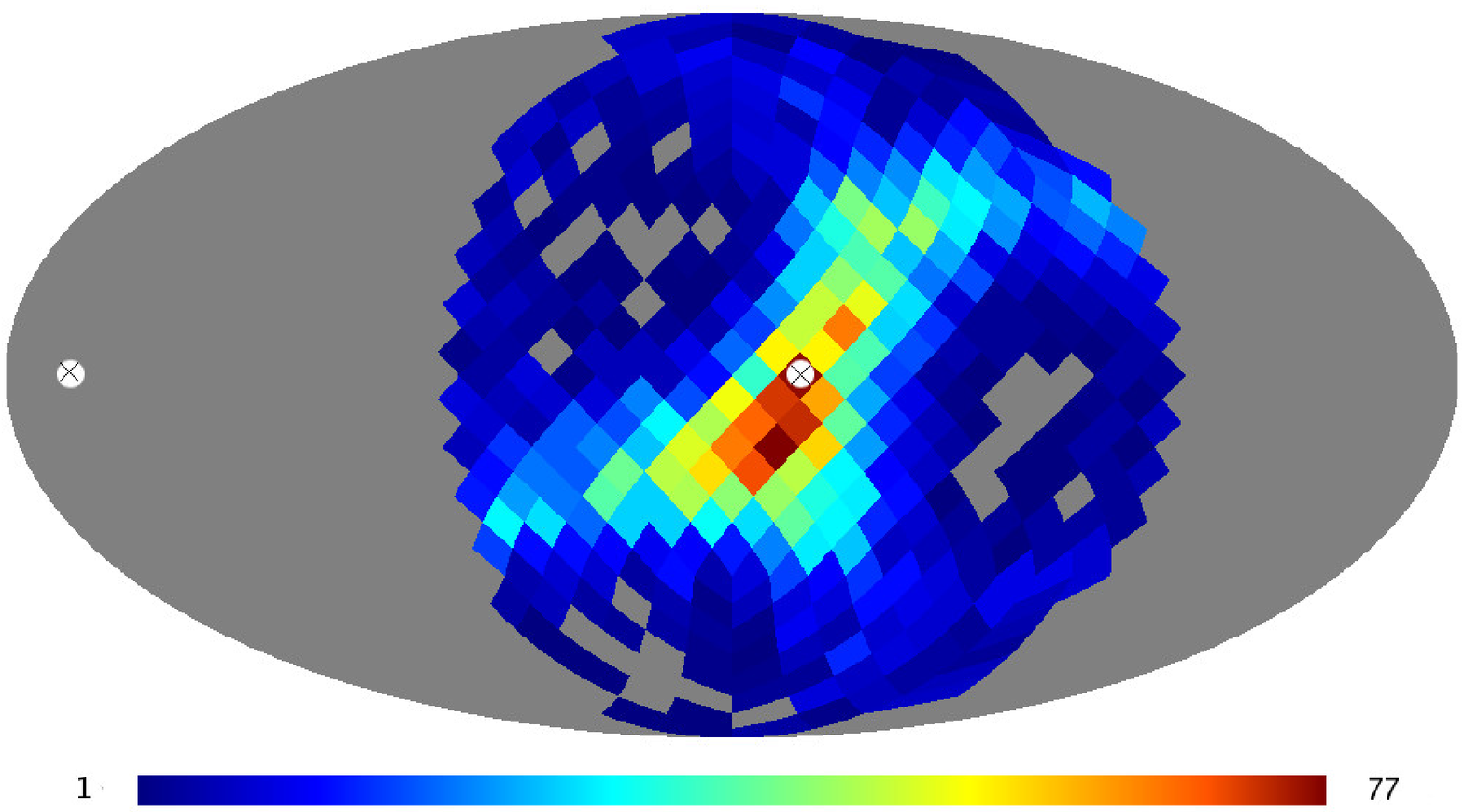}
 \caption{Preferred axis of the quadrupole (top panel) and the
   octopole (bottom panel) for $\Pred$. The colour coding counts the number of
   MC samples whose axis came to lie on the respective pixel.
   The axis of the quadrupole aligns with the axis of evil within our
   measurement precision, whereas the axis of the octopole does not.} 
 \label{fig:axes_Pred}
\end{figure}

\begin{figure}
 \centering
 \includegraphics[scale=0.4]{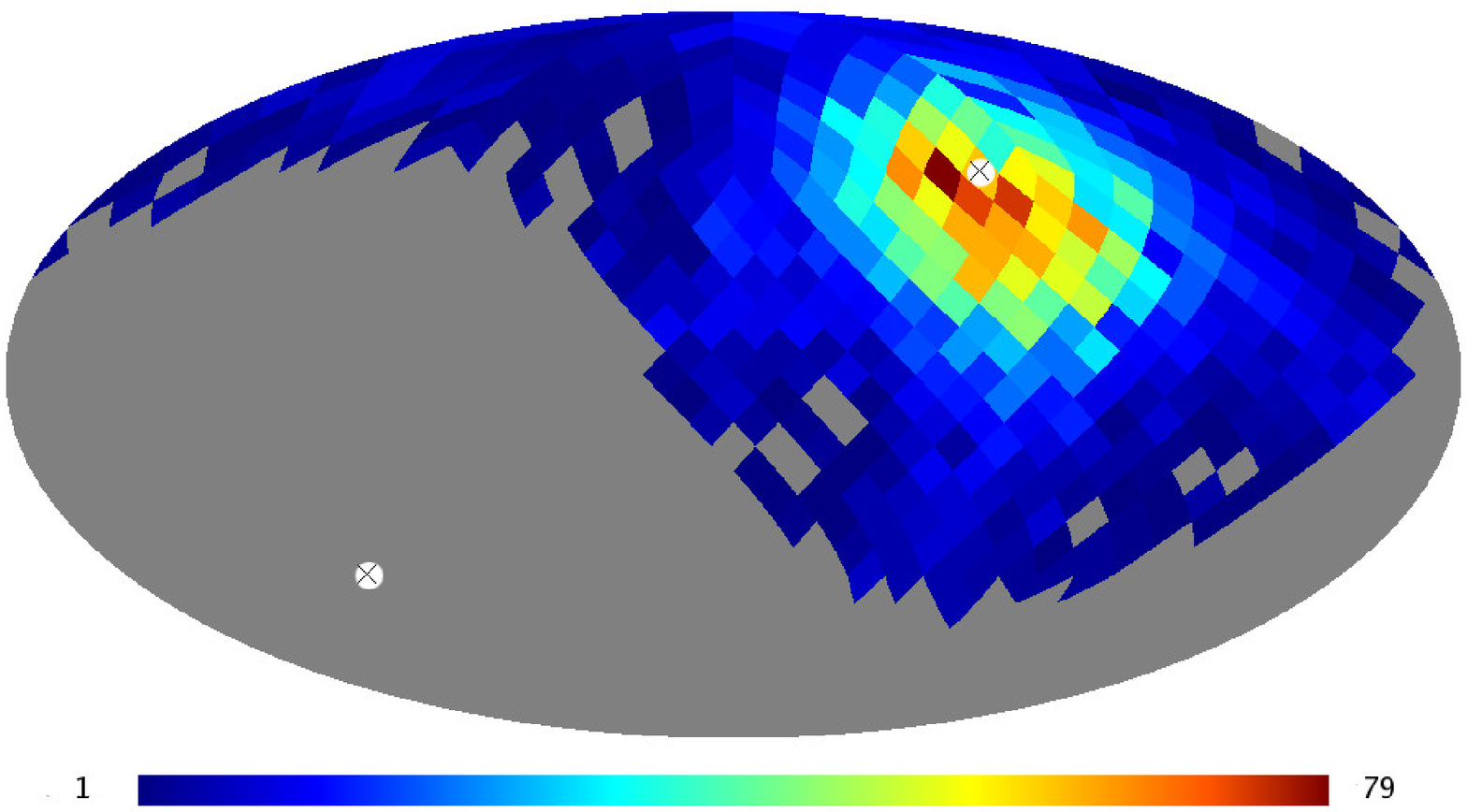}
 \includegraphics[scale=0.4]{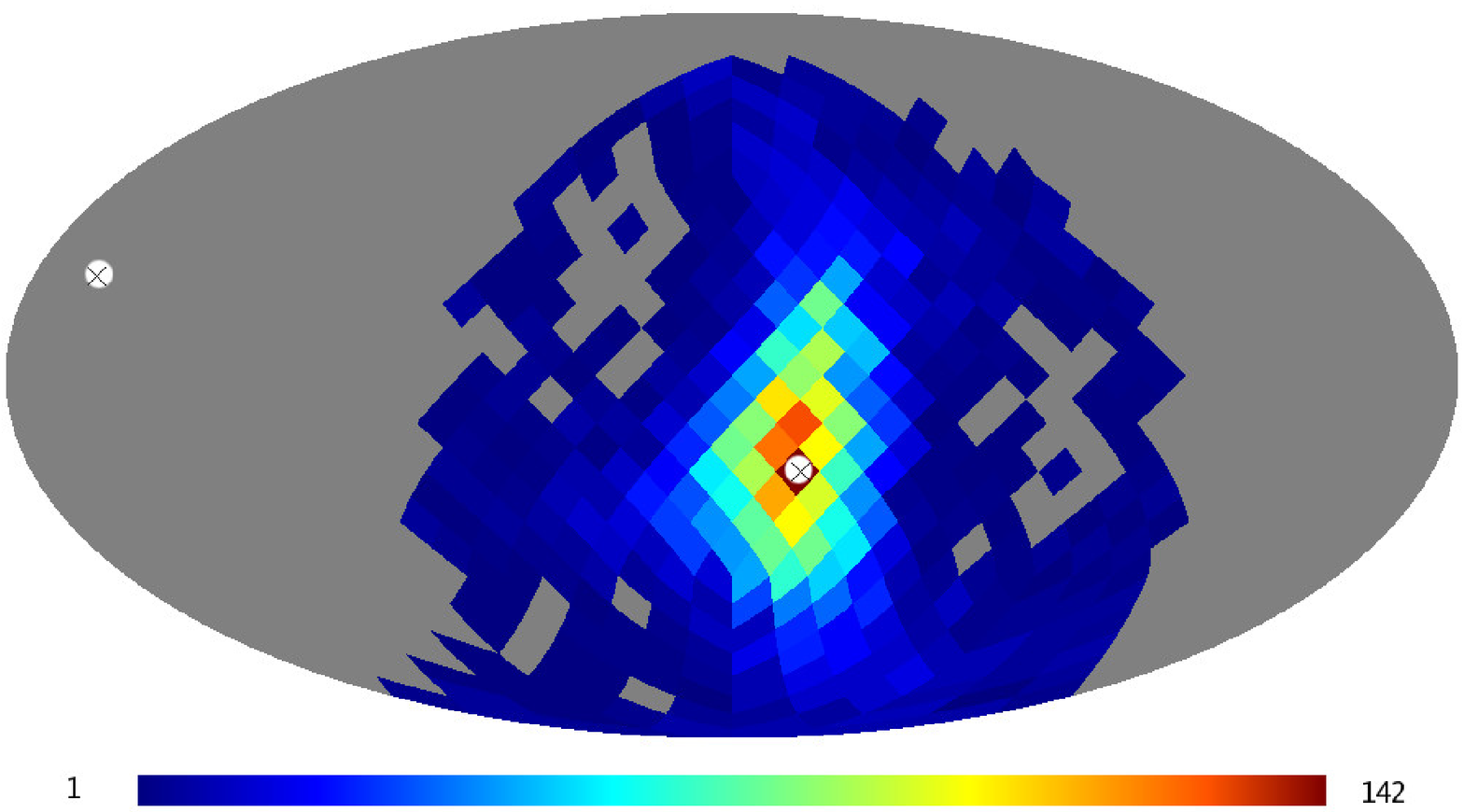}
 \caption{Preferred axis of the quadrupole (top panel) and the
   octopole (bottom panel) for $\Tknown$ and thus for $\Prec$. The
   axis of the quadrupole aligns with the axis of evil within our 
   measurement precision, whereas the axis of the octopole does not.}
 \label{fig:axes_Tknown}
\end{figure}

\begin{figure}
 \centering
 \includegraphics[scale=0.4]{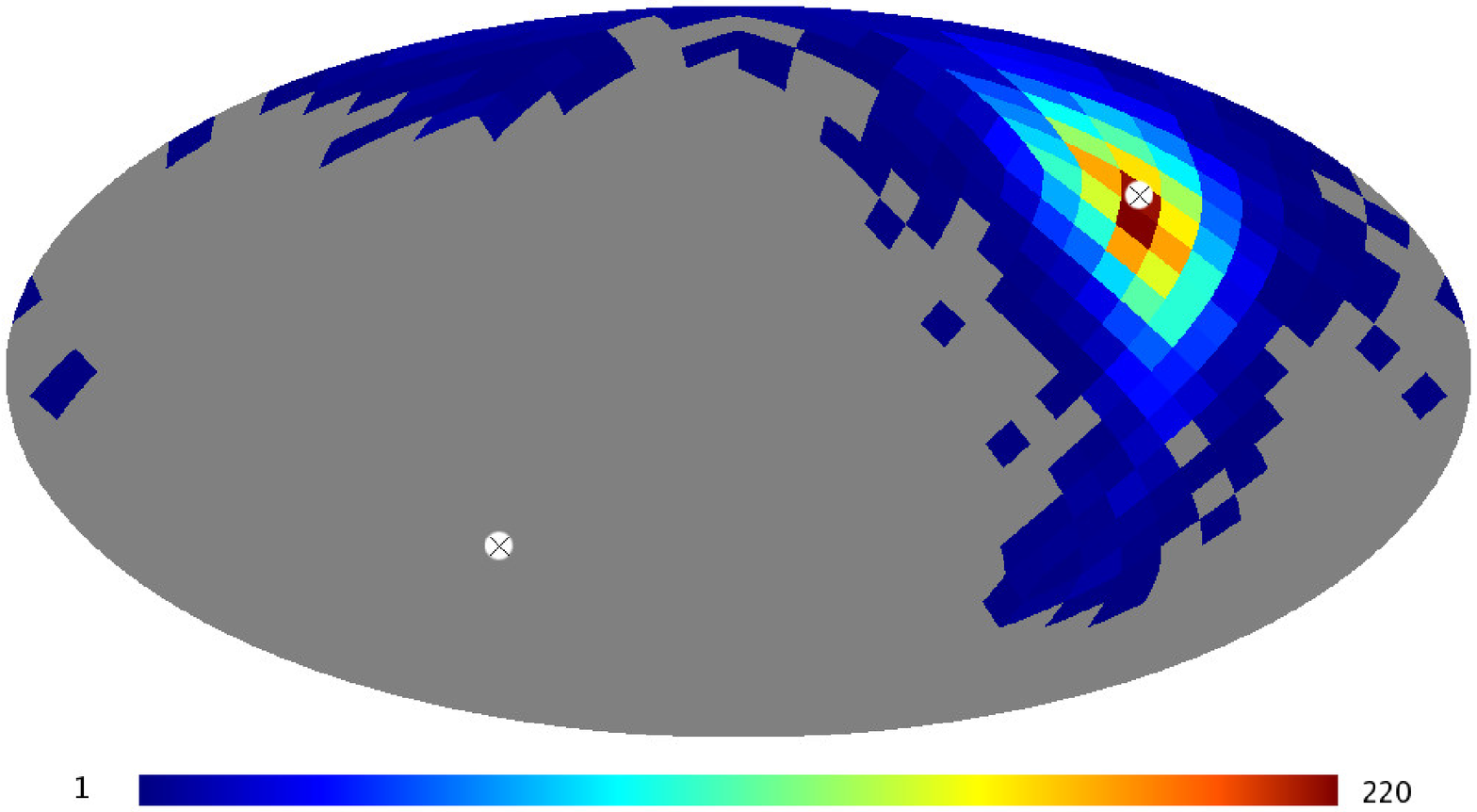}
 \includegraphics[scale=0.4]{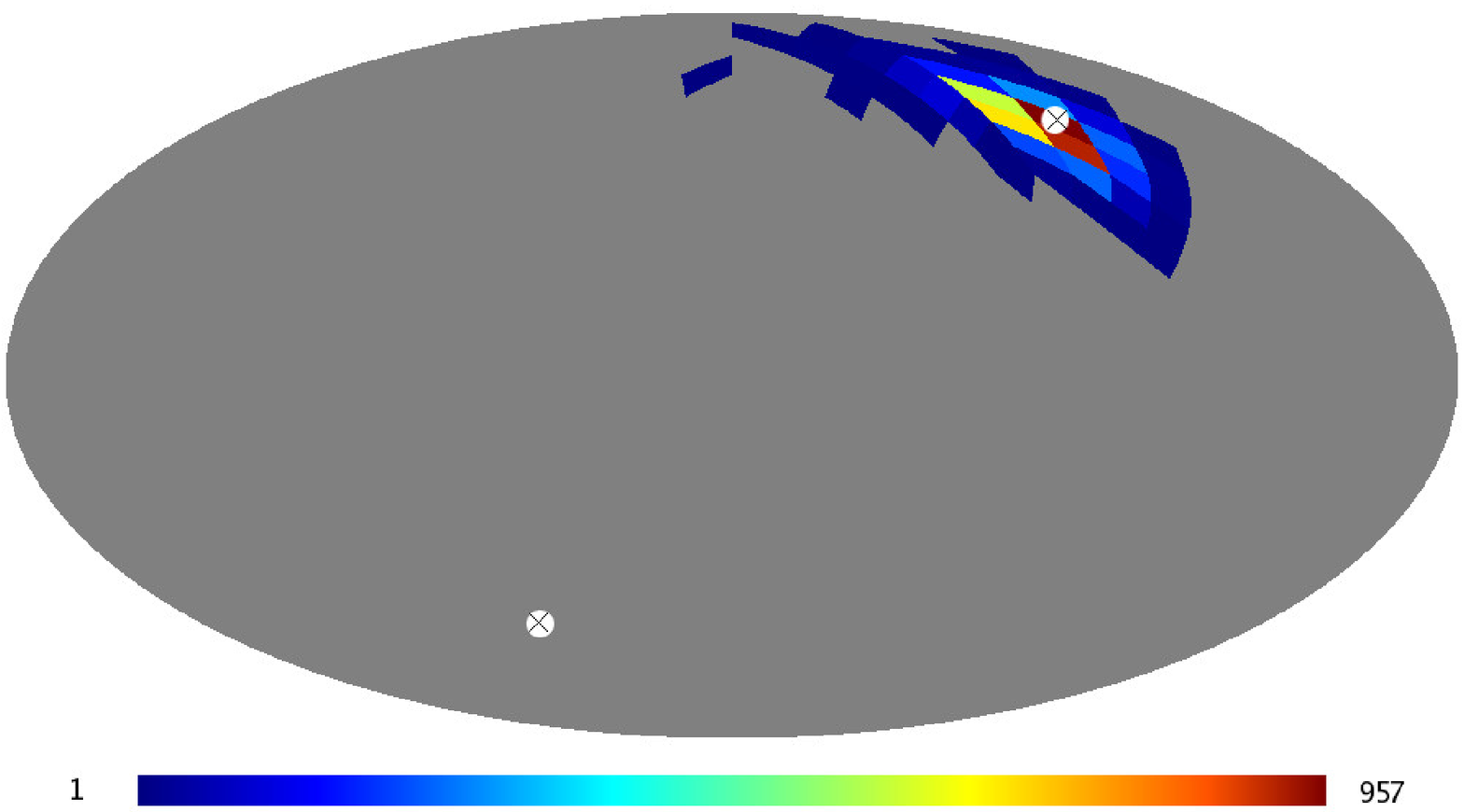}
 \caption{Preferred axis of the quadrupole (top panel) and the
   octopole (bottom panel) for $\Tred$. The axes of the quadrupole and
   the octopole both align with the axis of evil
   within our measurement precision.}
 \label{fig:axes_Tred}
\end{figure}

We plot the axes and their uncertainties for the different maps in
Figs \ref{fig:axes_Pknown} -- \ref{fig:axes_Tred}. 
Both ends of every axis are marked by a cross
in the maps, and the colour coding counts how many times the preferred
axis came to lie on the respective pixels in 5000 MC samples.
 
All axes and their standard deviations 
$\sigma$, which we obtained from the MC simulations, are
summarised in Table \ref{tab:axes}. 
For $\Pknown$, and thus the ILC map, we reproduce the results from
\cite{oliveira} within 
our measurement precision: the axes of the quadrupole and the octopole
of $\Pknown$ point in the same direction, which is roughly
$(l,b) \approx (-120^\circ,63^\circ)$, where $l$ and $b$ denote
Galactic longitude and latitude, respectively (\cite{oliveira} found
$(l,b) \approx (-110^\circ,60^\circ)$). For $\Tred$, again both axes  
point in the same direction as the axes of $\Pknown$ within our measurement
precision.  

For $\Pred$, the preferred axis of the quadrupole
has an angular distance to the average axis of the ILC map
of $37^\circ$. That means that the latter lies inside its $1\sigma$
region. The same holds for $\Tknown$ (and thus $\Prec$), 
for which the axis of the quadrupole has an angular distance
to the average axis of the ILC map of $34^\circ$. 
The axes of the octopole of $\Pred$ and $\Tknown$, though, do not
align with the axis of evil.
What can we learn from this result?
The significance of the alignment between the axes of the quadrupole and
octopole in the temperature map has been assessed extensively
in earlier works. In this work, we only look at the additional
information we obtain from the axes of $\Pred$. 
To this end, let us
take the preferred axis in the temperature map $\Tcmb$ as given, and
assume that the axes of $\Pred$ are distributed isotropically over the
sky and independently from each other. 
In Appendix \ref{sec:assess}, we work out the probability
for at least one of the axes of $\Pred$ being such that
the axis of the temperature map is included in the $1\sigma$ region
around it. This probability amounts to about 50 per cent, due to the large
$1\sigma$ regions we have. 

In order to assess whether the mask or the noise in the WMAP
polarization maps is the main source of uncertainty in the axes,
we have determined the uncertainty
with the amplitude of the noise covariance matrix rescaled
to 10 per cent of the original one. This yields an uncertainty
of about $20^\circ$ in the axes. We have done the same exercise for
the noise amplitude downscaled to 1 per cent of the original one,
which results in an uncertainty of $7^\circ-10^\circ$ in the
axes. This means that the noise is actually the main source of
uncertainty in our analysis rather than the mask.
Soon, the {\it Planck Surveyor} mission \citep{planck} will provide us
with polarization 
measurements that have a noise-level which is significantly
below the one in the WMAP data. The
main problem will then be the 
contamination of the polarization data by Galactic foregrounds. In the
WMAP polarization data, the foregrounds contribute about 20 per cent
to the diagonal of the noise covariance matrix $N_P$ in pixel space.
With {\it Planck}, we will be able to determine the foregrounds 
better than with WMAP, due to the broader frequency range covered by {\it
  Planck}. If we 
assume that the covariance due to residual foregrounds for {\it
  Planck} will be between 5 and 50 per cent of the one for WMAP, we
will get the uncertainty on the axes down to about $10^\circ$ and
$20^\circ$, respectively. With this, we will have a powerful test to
probe the axis of evil in polarization.

\section{Conclusions}\label{sec:conclusions}

In the last few years, a preferred axis has
been found in the CMB temperature map, posing a challenge to the
cosmological principle. This so-called {\it axis of evil} denotes the unusual
alignment of the preferred axes of the quadrupole and the octopole in
the temperature map.

In this work, we have split the CMB temperature and polarization maps
from WMAP into a part correlated with the respective other map, and an
uncorrelated part. If the axis of evil were due to some preferred
direction intrinsic to the geometry of the primordial Universe, we
would expect its signature to be present in all four of these maps, 
although this is not true for all theoretical models creating an axis
in the temperature map.
In particular, the part of the polarization map which is
uncorrelated with the temperature map serves as a statistically
independent probe of the axis of evil.

In order to reduce the noise contained in the polarization maps, we
have Wiener filtered the maps before computing the axes.
We have then determined the preferred axes of the
quadrupole and the octopole for the four maps. In order to assess the
uncertainty in the axes coming from the mask, detector noise and
residual foregrounds in the polarization maps, we have run MC
simulations conditional on the observational data.

For the part
of the polarization map which is correlated with the temperature map,
$\Pknown$, 
we find that the axes of quadrupole and octopole point in the same
direction, confirming earlier results by \cite{oliveira}.
The part of the temperature map which is uncorrelated with the
polarization map, $\Tred$, exhibits the same alignment of the axes within our
measurement precision.  

For the part of the polarization map which is uncorrelated with the
temperature map, $\Pred$, we find that only the axis of the quadrupole 
aligns with the axis of evil, whereas the axis of the octopole does
not. The same holds for the correlated part of the temperature map, $\Tknown$.
We have computed the probability that a rough alignment with
the axis of evil, as we find it for the axis of the quadrupole of
$\Pred$, happens by chance if the axes are distributed
isotropically. This probability amounts to 50 per cent for currently
available polarization data, due to
the large uncertainties in the axes. We are thus looking
forward to redoing this analysis with polarization maps from {\it Planck},
which will yield much more significant results.
Of course, similar analyses can be carried out for all other anomalies
that have been found in the CMB temperature maps.  
Note that, instead of working in pixel space as we have done, one
could implement the analysis in spherical harmonics space, which
would help to separate the E modes we are working with from
contamination by B modes.

The approach we have chosen here is a phenomenological approach,
since in principle one should take into account that
different models causing anomalies in the temperature map predict
different signatures in the polarization map.
Thus, for a more thorough analysis, one would need to
consider particular models of the primordial Universe creating
anomalies in the temperature maps, and compute the statistical
properties of the uncorrelated polarization map for these. This can be
done by modifying a Boltzmann code such as {\small CMBEASY} or by
simulations as in \cite{hu_polarization}. One can then try to find these
predicted signatures in the uncorrelated polarization map
via Bayesian model selection. Such an analysis would
truly go beyond the usual a posteriori analysis of finding anomalies in the
temperature map, since we would use an actual model to make
predictions for the uncorrelated polarization map and then compare
these predictions with observations. We leave this promising analysis for
future work.

\section*{acknowledgements}

The authors would like to thank Simon White for a suggestion that greatly
improved this paper. We further thank Thomas Riller, Carlos
Hernandez-Monteagudo, Christoph R{\"a}th, Anthony Banday, and the
anonymous referee for useful discussions and comments; Martin
Reinecke, Mariapaola Bottino, and Andr{\'e} Waelkens for their
help with {\small HEALPix}, and Claudia Scoccola for reading the manuscript. 
We acknowledge the use of the {\small HEALPix} package \citep{healpix}
, the {\it Legacy Archive for Microwave Background Data Analysis}
(LAMBDA), and {\small CMBEASY} \citep{cmbeasy}. 

\bibliographystyle{mn2e}
\bibliography{bibl.bib}

\begin{appendix}

\section{Proof of vanishing correlation between $\Tknown$ and
  $\Tred$}\label{sec:T_uncorr}

We now prove that the two maps $\Tknown$ and $\Tred$, into which we split the
temperature map $\Tcmb$ in section \ref{sec:split_t}, are indeed
uncorrelated. To this end, let us write
\bea \nonumber
\Tknown &=& S_{T,P}\, S_P^{-1} \Prec\\ \nonumber
&=& S_{T,P} S_P^{-1} (S_P^{-1} + \mask^\dagger N_P^{-1} \mask)^{-1}
\mask^\dagger N_P^{-1} \Pobs \\ \nonumber
&=& S_{T,P} (1 + \mask^\dagger N_P^{-1} \mask S_P)^{-1}
\mask^\dagger N_P^{-1} \Pobs \\ \nonumber
&=& S_{T,P} \mask^\dagger(1 + N_P^{-1} \mask S_P \mask^\dagger)^{-1} N_P^{-1} \Pobs 
\\
&=& S_{T,P} \mask^\dagger(N_P + \mask S_P \mask^\dagger)^{-1} \Pobs \,,
\label{T_known_3}
\eea
where we have inserted $\Prec$ from eq. (\ref{wiener_P}) in the first
step. The third step can be easily verified by 
using the geometric series for 
$\left(1 + \mask^\dagger N_P^{-1}\mask S_P \right)^{-1}\mask^\dagger$, which
has a convergence radius of 1, and is thus valid for $|\mask^\dagger
N_P^{-1}\mask S_P| < 1$. In our case, this holds because our polarization data
are noise-dominated.\footnote{By adding a small $\epsilon$-term to the
response $\mask$, and thus making it invertible, the third step also
holds generally.}

We will soon see that we need the following covariance matrices in the
derivation: 
\be
\langle \Pobs \Pobs^\dagger \rangle_{\pr(\Pobs \,|\, p)}
= \mask S_P \mask^\dagger + N_P\,, 
\label{C_p}
\ee
where we have assumed that $\Pcmb$ is uncorrelated with $\Pdet$ and $\Pfg$,
and we have inserted the definition of $N_P$, eq. (\ref{N_P}).

Since we neglect the detector noise and residual foregrounds in the
temperature data, we obtain for the
covariance between temperature and polarization data
\bea \nonumber
\langle \Tobs \Pobs^\dagger
\rangle_{\pr(\Tobs,\Pobs \,|\, p)} &=& \langle \Tcmb \Pcmb^\dagger
\rangle_{\pr(\Tcmb,\Pcmb \,|\, p)} \mask^\dagger \\ 
&\equiv& S_{T,P} \mask^\dagger\,,
\label{C_pt}
\eea
where we have assumed that detector noise and residual foregrounds in the
polarization map are uncorrelated with the CMB temperature map.

Let us now look at
\bea \nonumber
&& \langle \Tred \Tknown^\dagger \rangle_{\pr(\Tobs,\Pobs \,|\, p)} \\ \nonumber
&=& \langle \Tobs \Tknown^\dagger \rangle_{\pr(\Tobs,\Pobs \,|\, p)} - \langle
\Tknown \Tknown^\dagger 
\rangle_{\pr(\Tobs,\Pobs \,|\, p)} \\ \nonumber
&=& \langle \Tobs \Pobs^\dagger \rangle_{\pr(\Tobs,\Pobs \,|\, p)} (N_P +
\mask S_P \mask^\dagger)^{-1} \mask S_{P,T} \\ \nonumber
&& - \, S_{T,P} \mask^\dagger (N_P + \mask S_P \mask^\dagger)^{-1}
\langle \Pobs \Pobs^\dagger \rangle_{\pr(\Pobs \,|\, p)} \\ \nonumber
&& \;\;\;\, (N_P + \mask S_P \mask^\dagger)^{-1} \mask S_{P,T} \\ \nonumber
&=& S_{T,P} \mask^\dagger (N_P +
\mask S_P \mask^\dagger)^{-1} \mask S_{P,T} \\ \nonumber
&& - \, S_{T,P} \mask^\dagger (N_P + \mask S_P \mask^\dagger)^{-1}
(N_P + \mask S_P \mask^\dagger) \\ \nonumber
&& \;\;\;\, (N_P + \mask S_P \mask^\dagger)^{-1} \mask S_{P,T} \\ \nonumber
&=& S_{T,P} \mask^\dagger (N_P +
\mask S_P \mask^\dagger)^{-1} \mask S_{P,T} \\ \nonumber
&& - \, S_{T,P} \mask^\dagger (N_P + \mask S_P \mask^\dagger)^{-1}
\mask S_{P,T} \\ \nonumber
&=& 0\,,
\eea
where we have inserted eqs (\ref{T_known_3}), (\ref{C_p}), and (\ref{C_pt}).
QED

\section{Proof of vanishing correlation between $\Pknown$ and
  $\Pred$}\label{sec:P_uncorr}  

For the splitting of the polarization map, we first prove that the
unfiltered uncorrelated map defined in eq. (\ref{def_Predraw}),
$\Predraw$, is uncorrelated with $\Pknown$:  
\bea \nonumber
&& \langle \Predraw \Pknown^\dagger \rangle_{\pr(\Tobs,\Pobs \,|\, p)} \\
\nonumber 
&=& \langle \Pobs \Pknown^\dagger \rangle_{\pr(\Tobs,\Pobs \,|\, p)} - 
\mask \langle \Pknown \Pknown^\dagger \rangle_{\pr(\Tobs,\Pobs \,|\, p)} \\
\nonumber   
&=& \langle \Pobs \Tobs^\dagger \rangle_{\pr(\Tobs,\Pobs \,|\, p)} S_T^{-1}
S_{T,P} \\ \nonumber
&& - \, \mask S_{P,T} S_T^{-1} \langle \Tobs \Tobs^\dagger
\rangle_{\pr(\Tobs,\Pobs \,|\, p)} S_T^{-1} S_{T,P} \\ \nonumber   
&=& \mask S_{P,T} S_T^{-1} S_{T,P}  - \mask S_{P,T} S_T^{-1} S_T S_T^{-1}
S_{T,P} \\ \nonumber    
&=& \mask S_{P,T} S_T^{-1} S_{T,P}  - \mask S_{P,T} S_T^{-1} S_{T,P} \\
&=& 0\,.
\eea
From the above, we readily obtain that also the Wiener filtered uncorrelated
map, 
\bea \nonumber
\Pred &=& [(S_P - S_{P,T} S_T^{-1} S_{T,P})^{-1} + \mask^\dagger N_P^{-1}
  \mask]^{-1} \\ \nonumber
&& \mask^\dagger N_P^{-1}\Predraw,
\eea 
is uncorrelated with $\Pknown$:
\bea \nonumber
&& \langle \Pred \Pknown^\dagger \rangle_{\pr(\Tobs,\Pobs \,|\, p)} \\ \nonumber
&=& 
[(S_P - S_{P,T} S_T^{-1} S_{T,P})^{-1} + \mask^\dagger N_P^{-1}
  \mask]^{-1} \\ \nonumber
&& \mask^\dagger N_P^{-1}
\langle \Predraw \Pknown^\dagger \rangle_{\pr(\Tobs,\Pobs \,|\, p)} \\
&=& 0
\eea
QED

\section{Probability for chance alignment in an isotropic
  universe}\label{sec:assess} 

We would like to assess whether the rough alignment of the axis of the
quadrupole in $\Pred$ actually provides us with some information about
the axis of evil. We therefore compute the probability
for at least one of the axes of $\Pred$ aligning with the axis of the
temperature map in an isotropic universe.
To this end, let us
take the preferred axis in the temperature map $\Tcmb$ as given, and
assume that the axes of $\Pred$ are distributed isotropically over the
sky and independently from each other. 
We then work out the probability
for at least one of the axes of $\Pred$ being such that
the axis of the temperature map is included in the $1\sigma$ region
around it. 

For simplicity, we assume that the the $1\sigma$ regions are
symmetric circles around the axes, with radius $\sigma \approx
45^\circ$ for both the quadrupole and the octopole. The solid angle $A$
spanned by such a $1\sigma$ region is well approximated by $A \approx
\pi \sigma^2$.\footnote{This flat-sky approximation differs from the
  actual value of the solid angle by 6 per cent.}
The probability of at least one of the $1\sigma$ regions hitting the
axis of evil is just the solid angle spanned by the two $1\sigma$ regions
divided by the solid angle of the hemisphere, $2\pi$. However, the
solid angle spanned by the two $1\sigma$ regions
depends on the overlap $B$ between them, it is $2A-B$ to avoid double
counting of the overlapping area.
Given the angular separation $\alpha$ between the axes of the quadrupole and
the octopole, the overlap can be computed as follows:
\be
B(\alpha) = 2\left[\sigma^2 \arccos\left(\frac{\alpha}{2\sigma}\right)
  - \frac{\alpha}{2} \sqrt{\sigma^2 - \frac{\alpha^2}{4}}\right]\,,
\ee
which can be derived from the geometry of the problem in
flat-sky approximation. 
We marginalise the
hitting probability over the overlap $B(\alpha)$, using the fact that
$\alpha$ is distributed as $\pr(\alpha) = \sin(\alpha)$ \citep{oliveira}:
\bea \nonumber
\pr({\rm hit}) &=& \int_{\alpha = 0}^{\pi/2} \pr({\rm hit} \,|\,
B(\alpha)) \pr(\alpha) \,d\alpha \\
&=& \int_{\alpha = 0}^{\pi/2} \frac{2A-B(\alpha)}{2\pi} \sin(\alpha) \,d\alpha
\approx 50 \%\,.
\eea

\end{appendix}

\label{lastpage}

\end{document}